\documentclass[lettersize,journal]{IEEEtran}
\usepackage[cmex10]{amsmath}
\usepackage{amsfonts}
\usepackage{amssymb}
\usepackage{algorithmic}
\usepackage{algorithm}
\usepackage{array}

\usepackage{textcomp}
\usepackage{stfloats}
\usepackage{url}
\usepackage{verbatim}
\usepackage{graphicx}
\usepackage{cite}

\usepackage{soul}
\usepackage[dvipsnames]{xcolor}
\sethlcolor{CornflowerBlue}

\usepackage{multirow}

\usepackage[tight,footnotesize]{subfigure}
\usepackage{subfiles}

\newtheorem{proposition}{Proposition}
\newtheorem{remark}{Remark}

\begin{document}

\title{\hspace{0cm}Rate-Splitting Multiple Access:\\ Finite Constellations, Receiver Design,\\ and SIC-free Implementation}

\author{
    \IEEEauthorblockN{
        Sibo~Zhang,
        ~Bruno~Clerckx,~\IEEEmembership{Fellow,~IEEE,}
        ~David~Vargas,
        ~Oliver~Haffenden,
        ~Andrew~Murphy}
    \thanks{This work was supported in part by the UK Engineering and Physical Sciences Research Council, Industrial Case award number 210163.}
    \thanks{S. Zhang is with the Department of Electrical and Electronic Engineering at Imperial College London, London SW7 2AZ, UK and BBC Research and Development, The Lighthouse, White City Place, 201 Wood Lane, London, W12 7TQ, U.K. (e-mail: sibo.zhang19@imperial.ac.uk).}
    \thanks{B. Clerckx is with the Department of Electrical and Electronic Engineering at Imperial College London, London SW7 2AZ, UK (e-mail: b.clerckx@imperial.ac.uk).}
    \thanks{D. Vargas, O. Haffenden and A. Murphy are with BBC Research and Development, The Lighthouse, White City Place, 201 Wood Lane, London, W12 7TQ, U.K. (e-mail: david.vargas@bbc.co.uk; oliver.haffenden@bbc.co.uk; andrew.murphy@bbc.co.uk).} 
    }



\maketitle

\begin{abstract}
Rate-Splitting Multiple Access (RSMA) has emerged as a novel multiple access technique that enlarges the achievable rate region of Multiple-Input Multiple-Output (MIMO) broadcast channels with linear precoding. In this work, we jointly address three practical but fundamental questions: (1) How to exploit the benefit of RSMA under finite constellations? (2) What are the potential and promising ways to implement RSMA receivers? (3) Can RSMA still retain its superiority in the absence of successive interference cancellers (SIC)? To address these concerns, we first propose low-complexity precoder designs by taking finite constellations into account and show that the potential of RSMA is better achieved with such designs than those assuming Gaussian signalling. We then consider some practical receiver designs that can be applied to RSMA. We notice that these receiver designs follow one of two principles: (1) SIC: cancelling upper layer signals before decoding the lower layer and (2) non-SIC: treating upper layer signals as noise when decoding the lower layer. In light of this, we propose to alter the system design according to the receiver category. Through link-level simulations, the effectiveness of the proposed power allocation and receiver designs are verified. More importantly, we show that it is possible to preserve the superiority of RSMA over Spatial Domain Multiple Access (SDMA), including SDMA with advanced receivers, even without SIC at the receivers. Those results therefore open the door to competitive implementable RSMA strategies for 6G and beyond communications.
\end{abstract}


\section{Introduction}
\IEEEPARstart{T}{he} key goal of wireless communication systems is to achieve high data rates and spectral efficiencies. Mobile broadband standards, such as 4G and 5G, have converged to the use of linearly precoded Multi-User Multiple-Input Multiple-Output (MU-MIMO) technologies to support a large number of users with high data rates. Using beamforming, MU-MIMO multiplexes signals intended for different users in the spatial domain (hence also called Spatial Domain Multiple Access (SDMA)) and significantly improves the spectrum efficiency. To meet the foreseen requirement on even higher data rate and massive connectivity \cite{Tataria}, research attention has been drawn to novel multiple access techniques.

Among others, Rate-Splitting Multiple Access (RSMA) is favored as a candidate thanks to its numerous advantages in terms of efficiency, universality, flexibility, robustness, reliability, and low latency \cite{Clerckx_2, Mao_3}. With RSMA, in the simplest form, the transmitter first splits the unicast messages into common and private parts. The common parts are combined and encoded into a common stream while the private parts are encoded separately. The transmitter then utilizes linear precoding to send the common and private streams in a superposition manner. Each receiver decodes the common stream first and applies Successive Interference Cancellation (SIC) before decoding the desired private stream. Owing to the presence of both common and private streams, RSMA enables users to partially decode interference and partially treat interference as noise and hence is recognized as a powerful multiple access strategy. The idea of rate-splitting (RS) first emerged in the study of single antenna interference channel \cite{Carleial,Han}. \cite{Piovano,Hamdi,Mao} initiate the research on RSMA for multi-antenna broadcast channels by showing its superiority with partial Channel State Information at Transmitter (CSIT). \cite{Mao_2, Clerckx_1} show that even under perfect CSIT, RSMA generalizes and outperforms existing strategies, namely SDMA, Non-Orthogonal Multiple Access (NOMA), Orthogonal Multiple Access (OMA) and multicasting. Among others, 1-layer RSMA is the most attractive form of RSMA due to its low computation and hardware requirement \cite{Mao_2}. Concerning practical applications with low computational complexity requirements and latency requirement, low-complexity precoder designs for RSMA are studied in \cite{Clerckx_1,Dizdar_2,Lu,Hao,Flores,Zhou,Dai}. A common approach to obtain low-complexity precoders is to first design the precoder directions using conventional methods, and then decide the power allocation afterwards. In \mbox{\cite{Clerckx_1}}, a closed-form power allocation design for two-user downlink Multiple-Input Single-Output (MISO) with perfect CSIT is obtained. To tackle imperfect CSIT, \mbox{\cite{Dizdar_2}} proposes a closed-form power allocation considering mobility, while finite feedback is treated in \mbox{\cite{Lu, Hao}}. Designs for multiple receive antennas are considered in \cite{Flores}. In \cite{Zhou}, a low-complexity algorithm for spectral and energy efficiency trade-off problem is introduced. For millimeter wave systems, a low-complexity design for hybrid beamforming with finite feedback is studied in \cite{Dai}.

The aforementioned research on RSMA assumes Gaussian-distributed signalling from the transmitter and ideal SIC at the receiver for the ease of analysis. However, Gaussian signalling is impractical to realize and finite-alphabet constellations (such as Quadrature Amplitude Modulation (QAM) constellations) are commonly used. Very limited work studies the potential of RSMA with finite-alphabet constellations. In \cite{Salem}, constructive interference precoding with finite constellations is considered for RSMA, despite the fact that symbol-level precoding is practically difficult. In \cite{Dizdar,Dizdar_2,Yin,Chen_and_Mi}, with system designs that assume Gaussian signalling, Link-Level Simulation (LLS) performance of RSMA is evaluated under QAM constellations. However, it is difficult to exploit the potential of RSMA in practice without considering the actual constellations in use during the design process. System designs for RSMA that consider finite-alphabet constellations and are practically implementable are yet to be investigated.

For general wireless systems, research effort has been made on the theoretical analysis of finite-alphabet constellations. With constraints on constellations, the rate expressions, also called the Constellation Constrained (CC) rates \cite{Harshan}, significantly differ from that with Gaussian signalling and hence lead to different system designs. For example, uplink NOMA with finite constellations are studied in \cite{Harshan,Dong}. For MIMO downlink transmission, \cite{fin_const_1} proposes an iterative gradient-descent based algorithm to find the optimal precoder. Cooperative multi-cell transmission and multicast transmission are considered in \cite{Wu_2} and \cite{Wu_3}. Due to the non-trivial expressions, low-complexity designs considering finite-alphabet constellations for MU-MIMO are often hard to obtain.

Research on alternative RSMA receiver implementations, which potentially poses different complexity and performance, is very limited. \cite{Dizdar,Dizdar_2,Yin,Chen_and_Mi} only consider the very basic way of implementing SIC at the receiver side in the LLS. In \cite{loli2022modelbased}, the authors replace the de-modulation and cancellation modules in typical SIC receivers by deep neural networks and obtain performance improvement over basic SIC receivers under imperfect Channel State Information at Receiver (CSIR). \cite{joint_decoding,joint_decoding_2} analyze RSMA with joint decoding at the receiver from a purely information-theoretic perspective. However, implementation of joint decoding is very difficult in practice.

Another unresolved problem of RSMA lies in the necessity of SIC. Although it is essential to utilize SIC or joint decoding in information-theoretic analysis, the requirement of SIC poses a challenge in the practical implementation of RSMA because of its high complexity and error propagation issue\footnote{Despite being well know, complexity of SIC is quantified and demonstrated in Section IV. C and V. C, while error propagation effect is demonstrated Section V. C. Analysis on error propagation is also studied in \cite{Sena} and \cite{An}.}. The possibility of resolving the benefit of RSMA without SIC at receivers is completely undiscovered. 

In this work, we jointly address the three aforementioned practical design challenges of RSMA, namely system design with finite-alphabet constellation constraints, receiver implementation, and performance with non-SIC receivers. 
The contributions of this work are listed as follows:
\begin{enumerate}
    \item We derive the CC sum-rate expression and its approximation for RSMA with or without SIC. Assumptions on the precoders has been made such that the expressions are more tractable. Subject to actual constraints on the constellations in use, those rate expressions provide more accurate metrics for RSMA than those commonly used for Gaussian signalling.
    \item We simplify the problem of precoder design by fixing precoder directions and power distribution among the private streams. Based on the derived approximations of CC sum-rate expressions, we propose low-complexity methods on choosing the power allocation of common and private streams. Closed-form power allocation for optimizing the sum-rate assuming Gaussian signalling is also obtained as a bench mark.
    \item Despite being rarely studied in the literature, variations or substitutes of SIC are widely studied in standardization bodies \cite{DVB,NAICS,MUST}. In light of this, apart from the basic SIC receiver as in \cite{Dizdar,Dizdar_2,Yin,Chen_and_Mi}, we consider several receiver designs that can be applied with RSMA in real applications to provide more choices in terms of complexity and performance. Unlike works such as \cite{loli2022modelbased} where some components of a basic SIC receiver are improved, we focus on the overall architecture of the receiver and do not limit the design of any building blocks in it, that is, most of the de-mapper designs (such as maximum-likelihood de-mapper, max-log de-mapper and list sphere de-mapper) and decoder designs (such as belief propagation decoding and successive-cancellation list decoding for polar codes) can be adopted into the proposed receiver designs. The complexity of different receivers are quantified such that a complexity-performance trade-off with different receiver designs is revealed. We further notice that the receiver designs can be classified into SIC-type receivers and non-SIC-type receivers based on their principles and propose to alter the system design according to this classification. This also indicates that these receivers respectively pushes RSMA towards the two information theoretic limits, which are characterized by the CC sum-rate expressions assuming whether SIC is adopted or not.
    \item We evaluate RSMA with proposed power allocation methods and transceiver designs using Link-Level Simulations (LLS). We show that CC sum-rate oriented system designs lead to a significant performance improvement in practice in comparison with designs that assume Gaussian signalling and SDMA. We also show that, although non-SIC-type receivers lead to some performance loss compared to SIC-type receivers, they still preserve most of the superiority of RSMA over SDMA. \textbf{We reveal an important fact that it is possible to preserve the superiority of RSMA over SDMA, including SDMA with advanced receivers, even without SIC at the receivers.}
\end{enumerate}

\emph{Organization:} The remainder of this paper is organized as follows: Section II introduces the system model. In Section III, we propose three low-complexity precoder designs for RSMA. Section IV summarizes practical receiver implementations for RSMA. In Section V, numerical results are presented and analyzed. Finally Section VI concludes this paper.

\emph{Notations:}  Italic, bold lower-case, bold upper-case and calligraphic letters denote respectively scalars, vectors and matrices and sets. $\mathbf{I}$ denotes the identity matrix. $[\cdot]_{m,n}$ denotes the $(m,n)$-th entry of a matrix. $(\cdot)^T$, $(\cdot)^H$, $\text{tr}(\cdot)$ and $||\cdot||$ denote respectively the transpose, conjugate transpose, trace and Euclidean norm of the input entity. $|\cdot|$ denotes the absolute value if the argument is a scalar, or the cardinality if the argument is a set. $\times$ denotes the Cartesian product of two sets. $\mathsf{E}\{\cdot\}$ denotes the expectation. $\mathcal{CN}(\mu,\sigma^2)$ denotes circular symmetric complex Gaussian distribution with mean and variance to be $\mu$ and $\sigma^2$. 

\section{RSMA System Model}  
\begin{figure*}
      \centering
      \includegraphics[width=6in]{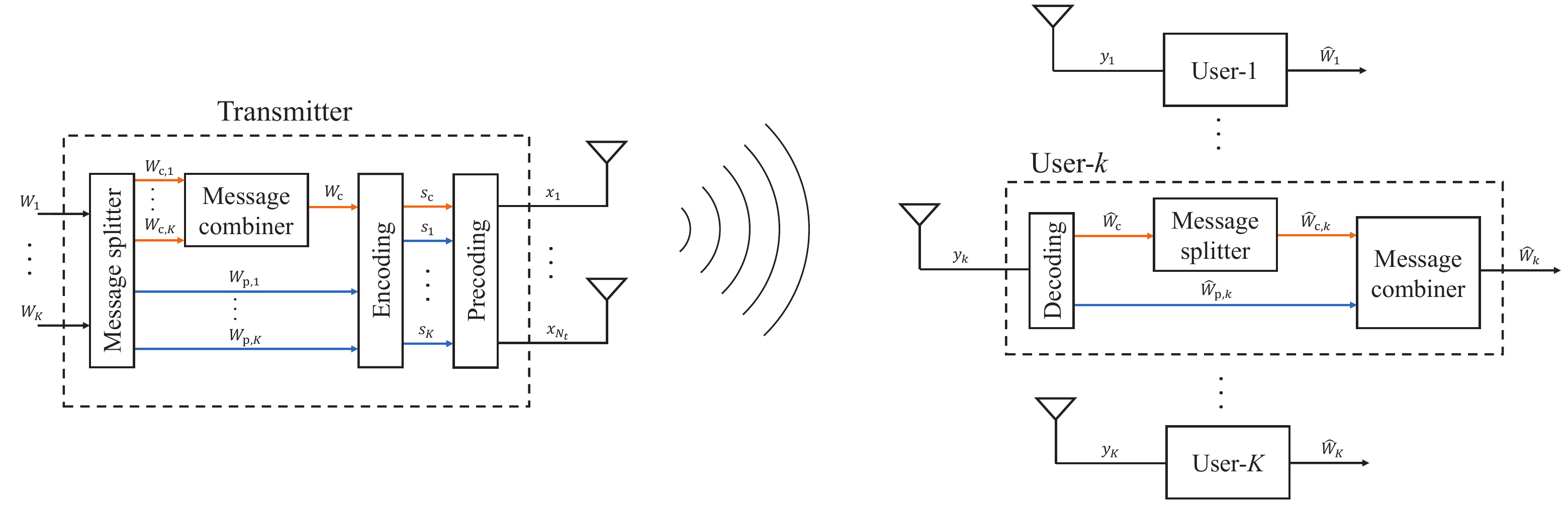}
      \caption{RSMA for multi-user MISO.}
      \label{RSMA}
\end{figure*}

We consider a multi-user MISO system, where a transmitter equipped with $N_t$ transmit antennas serves $K$ single-antenna unicast users under narrowband and fast fading channels. $\{W_1,...,W_K\}$ are the unicast messages intended for the $K$ users. We focus on underloaded scenarios, i.e. $N_t \geq K$. Perfect CSIT, CSIR and infinite block length are assumed. The received signal of user-$k$, $k \in \{1,...,K\}$, is expressed as
\begin{equation}
    y_k = \mathbf{h}_k^H \mathbf{x} + n_k ,
\end{equation}
where $\mathbf{h}_k \in \mathbb{C}^{N_t \times 1}$ is the channel vector for user-$k$, $\mathbf{x} \in \mathbb{C}^{N_t \times 1}$ is the transmitted signal vector and $n_k \sim \mathcal{CN}(0,1)$ is the additive white Gaussian noise at the receiver.

The transmitter utilizes linearly precoded single-layer RSMA \cite{Mao} whose key procedures are depicted in Fig. \ref{RSMA}.  At the transmitter side, $W_k$ is first split into common and private parts, $W_{\text{c},k}$ and $W_{\text{p},k}$, for all $k \in \{1,...,K\}$. The common parts from all the users are combined into a common message, $W_\text{c}$, and the private parts remain separate. Then the $K + 1$ messages are encoded separately into $K + 1$ data streams, $\mathbf{s} = [s_\text{c},s_1,...,s_K]^T \in \mathbb{C}^{(K+1) \times 1}$ and $\mathsf{E}\{\mathbf{s}\mathbf{s}^H\} = \mathbf{I}$. $\mathbf{s}$ is then precoded and transmitted. Upon reception, user-$k$ decodes the received signal into estimates of $W_{\text{c},k}$ and $W_{\text{p},k}$, $\widehat{W}_{\text{c},k}$ and $\widehat{W}_{\text{p},k}$, that are then combined to construct an estimate of the original unicast message intended for user-$k$, $\widehat{W}_{k}$. Precoder designs for RSMA are discussed in Section III. More details on transmitter and receiver implementation in the context of Bit-Interleaved Coded Modulation (BICM) are given in Section IV.

For practical considerations, we assume that power is uniformly distributed among the private streams and hence the transmit signal can be expressed as
\begin{equation}
    \mathbf{x} = \sqrt{P_\text{T}t}\Bar{\mathbf{p}}_\text{c} s_\text{c} + \sqrt{\frac{P_\text{T}(1-t)}{K}}\sum_{k=1}^{K} \Bar{\mathbf{p}}_k s_k,
\end{equation}
where $P_\text{T}$ is the total transmit power budget, $\{\Bar{\mathbf{p}}_\text{c},\Bar{\mathbf{p}}_1,...,\Bar{\mathbf{p}}_K\}$ are unit-norm vectors in $\mathbb{C}^{N_t \times 1}$ indicating direction of the precoders and $t\in [0,1]$ is the fraction of the power budget allocated to the common stream.

SDMA is a particular instance of RSMA when the common stream is turned off and hence $\{W_1,...,W_K\}$ are encoded into $K$ private data streams. In this work, precoders for SDMA are designed using Zero-Forcing (ZF) or Maximum Ratio Transmission (MRT) with uniform power allocation among the streams.

\section{Low-complexity precoder designs for RSMA}  
To obtain simple system designs for practical uses, we first decide the directions of the precoders, followed by the power allocation for common stream and private streams.

For the directions of the precoders, $\{\Bar{\mathbf{p}}_1,...,\Bar{\mathbf{p}}_K\}$ are designed using ZF precoding. Specifically, we define $\Bar{\mathbf{H}}  \triangleq [\Bar{\mathbf{h}}_1,\Bar{\mathbf{h}}_2,...,\Bar{\mathbf{h}}_K] = [\frac{\mathbf{h}_1}{||\mathbf{h}_1||}, \frac{\mathbf{h}_2}{||\mathbf{h}_2||},..., \frac{\mathbf{h}_K}{||\mathbf{h}_K||}]$. With ZF, we obtain $[\Bar{\mathbf{p}}_1,...,\Bar{\mathbf{p}}_K] = \Bar{\mathbf{H}}(\Bar{\mathbf{H}}^H \Bar{\mathbf{H}})^{-1}$. For $\Bar{\mathbf{p}}_\text{c}$, heuristically, we use solutions for the following multicast problem\footnote{Any other choice of $\Bar{\mathbf{p}}_\text{c}$ is valid here without affecting any other parts of the rest of the design.}
\begin{equation}
\begin{split}
    \mathcal{P}_1: \underset{\Bar{\mathbf{p}}_\text{c}}{\text{max}} \: \: & \underset{k}{\text{min}} |\Bar{\mathbf{h}}_k^H \Bar{\mathbf{p}}_\text{c}|^2\\ 
    \text{s.t.} \: \: \: \;
    & ||\Bar{\mathbf{p}}_\text{c}||^2 = 1.
\end{split}
\end{equation}
Solution to $\mathcal{P}_1$ for general cases has been addressed in \cite{Multicast_1, Multicast_2, Multicast_3}. Specifically, for 2-user cases, \mbox{\cite{Multicast_3}} provides the globally optimal $\Bar{\mathbf{p}}_\text{c}$ in closed form assuming that $|\Bar{\mathbf{h}}_1^H \Bar{\mathbf{p}}_\text{c}|^2 = |\Bar{\mathbf{h}}_2^H \Bar{\mathbf{p}}_\text{c}|^2$ is necessary for the solution. This solution is applied in this work for $K=2$. For $K\geq3$, dominant left singular vector of $\Bar{\mathbf{H}}$ is used as $\Bar{\mathbf{p}}_\text{c}$ for its low complexity but other designs, for example, those in \mbox{\cite{Multicast_1}} and \mbox{\cite{Multicast_2}}, are also valid.

With $\{\Bar{\mathbf{p}}_\text{c},\Bar{\mathbf{p}}_1,...,\Bar{\mathbf{p}}_K\}$ decided, $t$ is the only variable to be confirmed. We propose two methods on deciding $t$, namely optimizing sum-rate and optimizing CC sum-rate\footnote{To clarify, in this work, sum-rate implies that all the streams are transmitted using Gaussian signals, while CC sum-rate implies that the constellation is constrained to be a discrete and finite set.}.

\subsection{Sum-Rate Optimization}
The power allocation coefficient, $t$, is obtained by solving a sum-rate maximization problem. With the above choice of $\{\Bar{\mathbf{p}}_c,\Bar{\mathbf{p}}_1,...,\Bar{\mathbf{p}}_K\}$ and the assumption of Gaussian signalling, the sum-rate of RSMA can be written as
\begin{equation}\label{sum_rate}
\begin{split}
    R_{\text{sum}} = &\underset{k}{\text{min}} \: \text{log}_2 \Big( 1+\frac{P_\text{T}t|\mathbf{h}_k^H \Bar{\mathbf{p}}_\text{c}|^2}{\frac{P_\text{T}(1-t)}{K}|\mathbf{h}_k^H \Bar{\mathbf{p}}_k|^2 + 1}\Big)\\
    &+ \sum_{k=1}^{K} \text{log}_2\Big({1+\frac{P_\text{T}(1-t)}{K}|\mathbf{h}_k^H \Bar{\mathbf{p}}_k|^2}\Big),
\end{split}
\end{equation}
where we have used the fact that $|\mathbf{h}_k^H \Bar{\mathbf{p}}_i|^2 = 0, \forall i \neq k$. The first term in (\ref{sum_rate}) represents the achievable rate of the common stream such that it can be decoded by all the users. The second term represents the sum of the achievable rates of all the private streams.
It is trivial to see that
\begin{equation}
\begin{split}
       k'\triangleq&\text{arg} \underset{k}{\text{min}} \: \text{log}_2 \Big( 1+\frac{P_\text{T}t|\mathbf{h}_k^H \Bar{\mathbf{p}}_\text{c}|^2}{\frac{P_\text{T}(1-t)}{K}|\mathbf{h}_k^H \Bar{\mathbf{p}}_k|^2 + 1}\Big)\\
        =&
        \footnotemark
        \text{arg} \underset{k}{\text{min}} \: \frac{|\mathbf{h}_k^H \Bar{\mathbf{p}}_\text{c}|^2}{|\mathbf{h}_k^H \Bar{\mathbf{p}}_k|^2}
\end{split}
\end{equation}
\footnotetext{This equality holds if $|\Bar{\mathbf{h}}_1^H \Bar{\mathbf{p}}_\text{c}| = |\Bar{\mathbf{h}}_2^H \Bar{\mathbf{p}}_\text{c}| = ... = |\Bar{\mathbf{h}}_K^H \Bar{\mathbf{p}}_\text{c}|$ and holds asymptotically with $P \rightarrow \infty$ if otherwise.}and $k'$ depends on $\{\Bar{\mathbf{p}}_c,\Bar{\mathbf{p}}_1,...,\Bar{\mathbf{p}}_K\}$ but not $t$ and therefore can be pre-determined. 

We can then further simplify (\ref{sum_rate}) into
\begin{equation}\label{sum_rate_2}
\begin{split}
        R_{\text{sum}} =& \text{log}_2 \Big( 1 + P_\text{T}t|\mathbf{h}_{k'}^H \Bar{\mathbf{p}}_\text{c}|^2 +  \frac{P_\text{T}(1-t)}{K}|\mathbf{h}_{k'}^H \Bar{\mathbf{p}}_{k'}|^2 \Big)\\
        &+ \sum_{k=1,k\neq k'}^{K} \text{log}_2\Big({1+\frac{P_\text{T}(1-t)}{K}|\mathbf{h}_k^H \Bar{\mathbf{p}}_k|^2}\Big).
\end{split}
\end{equation}
(\ref{sum_rate_2}) is a concave function of $t$. Let $t^\star$ be the globally optimal $t$ that maximizes ({\ref{sum_rate_2}}). Using the first-order derivative, $t^\star$ is given by

\begin{equation}
    t^\star = 
    \left \{ \begin{array}{rcl}
    0 &\mbox{if} &-\sum_{i}\frac{a_i}{1+a_i}+\frac{b+a_{k'}}{1+a_k'} \leq 0\\
    1 &\mbox{if} &-\sum_{i}\frac{a_i}{1+a_i}+\frac{b+a_{k'}}{1+a_k'}>0 \mbox{ and } \\ & & -\sum_{i\neq k}a_i + \frac{b}{1+a_{k'}+b}\geq 0\\
    t_0 & & \hspace{-0.6cm} \mbox{otherwise},\\
    \end{array}\right.  
\end{equation}
where $a_k = \frac{P_{\text{T}}}{K} |\mathbf{h}_k^H \Bar{\mathbf{p}}_k|^2, \;k \in \{1,...,K\}$, $b = P_{\text{T}}|\mathbf{h}_{k'}^H \Bar{\mathbf{p}}_\text{c}|^2 - \frac{P_{\text{T}}}{K} |\mathbf{h}_{k'}^H \Bar{\mathbf{p}}_{k'}|^2$ and $t_0$ is the only real root of the following polynomial of $t$,
\begin{equation}
\begin{split}
   &\sum_{i=1,i\neq k'}^{K} a_i \prod_{j=1,j\neq i, k'}^{K} (t a_j - a_j - 1)(tb + a_{k'} + 1)\\
   &+ b \prod_{i=1, i\neq k'}^{K} (-ta_i + a_i + 1).
\end{split}
\end{equation} 

While the assumption of Gaussian signalling eases the analysis of the system performance, practical systems utilize finite constellations only. Next, two designs considering finite-alphabet constellations are introduced.

\subsection{Constellation-Constrained Sum-Rate Optimization}
In what follows, designing $t$ to optimize the achievable sum rate with finite-alphabet constellations, i.e. CC sum-rate, is addressed. We consider two cases: 

\begin{enumerate}
    \item all the users apply SIC receivers: each user can perfectly remove common stream signals before decoding the desired private stream, as long as the common stream is decodable by all the users;
    \item none of the users applies SIC receivers: each user decodes the private stream without any a-priori knowledge on the common message, i.e. treating the common stream signal as noise when decoding the desired private stream. Readers are referred to Section IV C for more discussion on implementing SIC-free receivers.
\end{enumerate}

\begin{figure*}
\hrule
\begin{flalign}\label{CC_sum_rate_SIC}
    R_{\text{CC},\text{sum}}^{\text{SIC}}
    = 
    &\log_2 |\mathcal{X}_\text{c}| - \frac{K}{\ln2} +\min_{k \in \{1,...,K\}}  \Bigg( -\frac{1}{|\mathcal{X}_\text{c}\times\mathcal{X}_k|} \sum_{m=1}^{|\mathcal{X}_\text{c}\times\mathcal{X}_k|} \mathsf{E}_n \Big\{ \log_2 \sum_{l=1}^{|\mathcal{X}_\text{c}\times\mathcal{X}_k|} \exp \Big( -\frac{|\mathbf{h}_k^H [\mathbf{p}_\text{c}\:\mathbf{p}_k] (\mathbf{s}_{k,m} - \mathbf{s}_{k,l}) + n|^2}{\sigma^2} \Big) \Big\}&&\nonumber\\
    & +\frac{1}{|\mathcal{X}_k|} \sum_{m=1}^{|\mathcal{X}_k|} \mathsf{E}_n \Big\{ \log_2 \sum_{l=1}^{|\mathcal{X}_k|} \exp \Big( -\frac{|\mathbf{h}_k^H \mathbf{p}_k(s_{k,m} - s_{k,l}) + n|^2}{\sigma^2} \Big) \Big\} \Bigg)&&\nonumber\\
    & + \sum_{k'=1}^{K} \log_2 |\mathcal{X}_{k'}| - \frac{1}{|\mathcal{X}_{k'}|} \sum_{m=1}^{|\mathcal{X}_{k'}|} \mathsf{E}_n \Big\{ \log_2 \sum_{l=1}^{|\mathcal{X}_{k'}|} \exp \Big( -\frac{|\mathbf{h}_{k'}^H \mathbf{p}_{k'}(s_{{k'},m} - s_{{k'},l}) + n|^2}{\sigma^2} \Big) \Big\}&&
\end{flalign}
    
\begin{flalign}\label{CC_sum_rate_approx_SIC}
    R_{\text{CC},\text{sum}}^{\text{SIC}}
    \approx 
    &\log_2 |\mathcal{X}_\text{c}| + \min_{k \in \{1,...,K\}}  \Bigg( -\frac{1}{|\mathcal{X}_\text{c}\times\mathcal{X}_k|} \sum_{m=1}^{|\mathcal{X}_\text{c}\times\mathcal{X}_k|} \log_2 \sum_{l=1}^{|\mathcal{X}_\text{c}\times\mathcal{X}_k|} \exp \Big( -\frac{|\mathbf{h}_k^H [\mathbf{p}_\text{c}\:\mathbf{p}_k] (\mathbf{s}_{k,m} - \mathbf{s}_{k,l})|^2}{2\sigma^2} \Big)&&\nonumber\\
    & +\frac{1}{|\mathcal{X}_k|} \sum_{m=1}^{|\mathcal{X}_k|} \log_2 \sum_{l=1}^{|\mathcal{X}_k|} \exp \Big( -\frac{|\mathbf{h}_k^H \mathbf{p}_k(s_{k,m} - s_{k,l})|^2}{2\sigma^2} \Big) \Bigg)&&\nonumber\\
    &+ \sum_{k'=1}^{K} \log_2 |\mathcal{X}_{k'}| - \frac{1}{|\mathcal{X}_{k'}|} \sum_{m=1}^{|\mathcal{X}_{k'}|} \log_2 \sum_{l=1}^{|\mathcal{X}_{k'}|} \exp \Big( -\frac{|\mathbf{h}_{k'}^H \mathbf{p}_{k'}(s_{k',m} - s_{k',l})|^2}{2\sigma^2} \Big)&&
\end{flalign}

\begin{flalign}\label{CC_sum_rate_no_SIC}
    R_{\text{CC},\text{sum}}^{\text{non-SIC}}
    = 
    & \log_2 |\mathcal{X}_\text{c}| + \min_{k \in \{1,...,K\}}  \Bigg( -\frac{1}{|\mathcal{X}_\text{c}\times\mathcal{X}_k|} \sum_{m=1}^{|\mathcal{X}_\text{c}\times\mathcal{X}_k|} \mathsf{E}_n \Big\{ \log_2 \sum_{l=1}^{|\mathcal{X}_\text{c}\times\mathcal{X}_k|} \exp \Big( -\frac{|\mathbf{h}_k^H [\mathbf{p}_\text{c}\:\mathbf{p}_k] (\mathbf{s}_{k,m} - \mathbf{s}_{k,l}) + n|^2}{\sigma^2} \Big) \Big\}\; \; \; \; \; \; \; \; \; \; \; \; \; \; \; &&\nonumber\\
    & +\frac{1}{|\mathcal{X}_k|} \sum_{m=1}^{|\mathcal{X}_k|} \mathsf{E}_n \Big\{ \log_2 \sum_{l=1}^{|\mathcal{X}_k|} \exp \Big( -\frac{|\mathbf{h}_k^H \mathbf{p}_k(s_{k,m} - s_{k,l}) + n|^2}{\sigma^2} \Big) \Big\} \Bigg) &&\nonumber\\
    &+\sum_{k'=1}^{K} \log_2 |\mathcal{X}_{k'}| - \frac{1}{|\mathcal{X}_\text{c}\times\mathcal{X}_{k'}|} \sum_{m=1}^{|\mathcal{X}_\text{c}\times\mathcal{X}_{k'}|} \mathsf{E}_n \Big\{ \log_2 \sum_{l=1}^{|\mathcal{X}_\text{c}\times\mathcal{X}_{k'}|} \exp \Big( -\frac{|\mathbf{h}_{k'}^H [\mathbf{p}_\text{c}\:\mathbf{p}_k] (\mathbf{s}_{k',m} - \mathbf{s}_{k',l}) + n|^2}{\sigma^2} \Big) \Big\} &&\nonumber\\
    &+ \frac{1}{|\mathcal{X}_\text{c}|} \sum_{m=1}^{|\mathcal{X}_\text{c}|} \mathsf{E}_n \Big\{ \log_2 \sum_{l=1}^{|\mathcal{X}_\text{c}|} \exp \Big( -\frac{|\mathbf{h}_{k'}^H \mathbf{p}_\text{c}(s_{\text{c},m} - s_{\text{c},l}) + n|^2}{\sigma^2} \Big) \Big\}&&
\end{flalign}

\begin{flalign}\label{CC_sum_rate_approx_no_SIC}
\raggedright
    R_{\text{CC},\text{sum}}^{\text{non-SIC}}
    \approx 
    &\log_2 |\mathcal{X}_\text{c}| + \min_{k \in \{1,...,K\}}  \Bigg( -\frac{1}{|\mathcal{X}_\text{c}\times\mathcal{X}_k|} \sum_{m=1}^{|\mathcal{X}_\text{c}\times\mathcal{X}_k|} \log_2 \sum_{l=1}^{|\mathcal{X}_\text{c}\times\mathcal{X}_k|} \exp \Big( -\frac{|\mathbf{h}_k^H [\mathbf{p}_\text{c}\:\mathbf{p}_k] (\mathbf{s}_{k,m} - \mathbf{s}_{k,l})|^2}{2\sigma^2} \Big)&&\nonumber\\ 
    &+\frac{1}{|\mathcal{X}_k|} \sum_{m=1}^{|\mathcal{X}_k|} \log_2 \sum_{l=1}^{|\mathcal{X}_k|} \exp \Big( -\frac{|\mathbf{h}_k^H \mathbf{p}_k(s_{k,m} - s_{k,l})|^2}{2\sigma^2} \Big) \Bigg)&&\nonumber\\ 
    & +\sum_{k'=1}^{K} \log_2 |\mathcal{X}_{k'}| - \frac{1}{|\mathcal{X}_\text{c}\times\mathcal{X}_{k'}|} \sum_{m=1}^{|\mathcal{X}_\text{c}\times\mathcal{X}_k'|} \log_2 \sum_{l=1}^{|\mathcal{X}_\text{c}\times\mathcal{X}_k'|} \exp \Big( -\frac{|\mathbf{h}_{k'}^H [\mathbf{p}_\text{c}\:\mathbf{p}_k] (\mathbf{s}_{k',m} - \mathbf{s}_{k',l})|^2}{2\sigma^2} \Big)&&\nonumber\\
    & +\frac{1}{|\mathcal{X}_\text{c}|} \sum_{m=1}^{|\mathcal{X}_\text{c}|} \log_2 \sum_{l=1}^{|\mathcal{X}_\text{c}|} \exp \Big( -\frac{|\mathbf{h}_{k'}^H \mathbf{p}_\text{c}(s_{\text{c},m} - s_{\text{c},l})|^2}{2\sigma^2} \Big)&&
\end{flalign}  
\hrule
\end{figure*}

\begin{proposition}\label{prop_1}
    The CC sum-rate of RSMA with SIC receivers is expressed as (\ref{CC_sum_rate_SIC}) and can be well approximated by (\ref{CC_sum_rate_approx_SIC}), where $\mathbf{p}_\text{c} = \sqrt{P_\text{T}t}\Bar{\mathbf{p}}_\text{c}$, $\mathbf{p}_k = \sqrt{\frac{P_\text{T}(1-t)}{K}}\Bar{\mathbf{p}}_k$, $\mathcal{X}_\text{c}$ is the finite size constellation set used for the common stream, $\mathcal{X}_k$ is the constellation set used for the private stream intended for the $k$-th user, $s_{k,m}$ is the $m$-th element in $\mathcal{X}_k$ and $\mathbf{s}_{k,m} \in \mathbb{C}^{2 \times 1}$ contains two entries drawn from the $m$-th element in $\mathcal{X}_\text{c}\times\mathcal{X}_k$.
\end{proposition}

\begin{IEEEproof}
    See Appendix A.
\end{IEEEproof}

\begin{proposition}\label{prop_2}
    The CC sum-rate of RSMA without SIC receivers is expressed as (\ref{CC_sum_rate_no_SIC}) and can be well approximated by (\ref{CC_sum_rate_approx_no_SIC}).
\end{proposition}

\begin{IEEEproof}
    See Appendix A.
\end{IEEEproof}

$t^\star$ is obtained by searching for the point in $[0,1]$ that maximizes either $R_{\text{CC},\text{sum}}^{\text{SIC}}$ or $R_{\text{CC},\text{sum}}^{\text{non-SIC}}$, for scenarios where users apply or not apply SIC respectively. 

\begin{figure}[!t]
      \centering
      \includegraphics[width=2.6in]{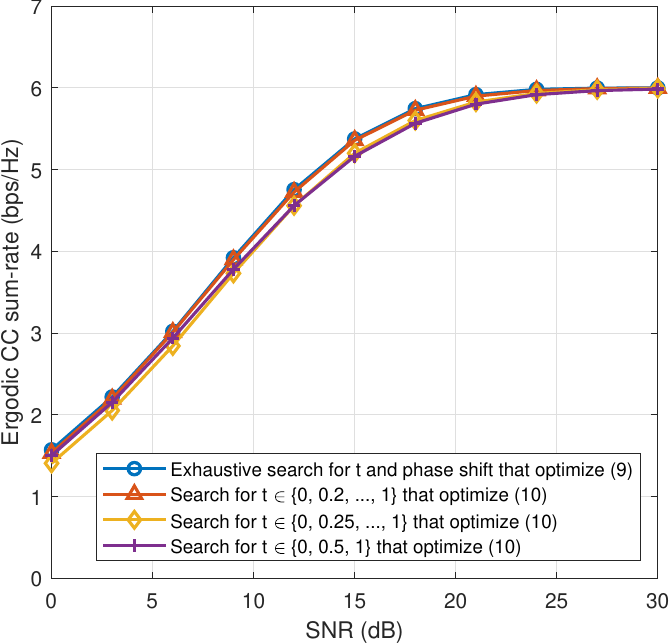}%
      \caption{Comparison of CC sum-rate for RSMA with SIC achieved by different methods of searching for $t^\star$ with $N_t=4$, $K=2$.}
      \label{search_space_SIC}
      \vspace{0.5cm}

      \centering
      \includegraphics[width=2.6in]{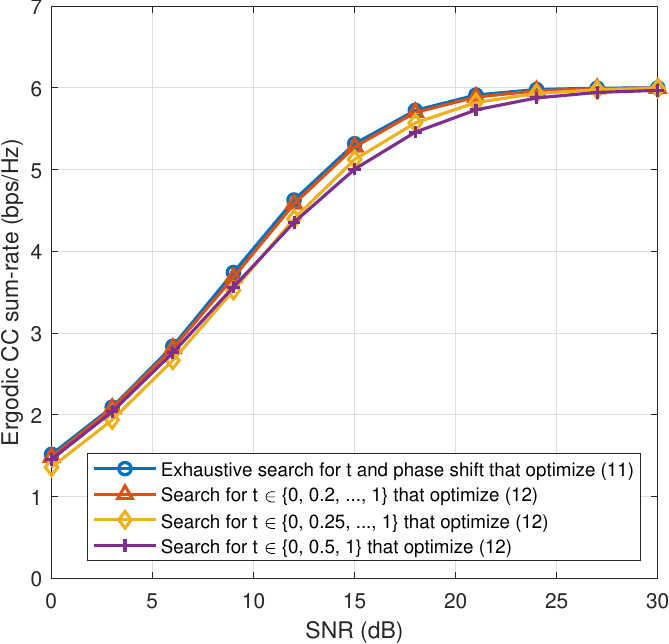}%
      \caption{Comparison of CC sum-rate for RSMA without SIC achieved by different methods of searching for $t^\star$ with $N_t=4$, $K=2$.}
      \label{search_space_non_SIC}
\end{figure}

\begin{remark}[Phase shift and search space]
Because of its circular symmetry, a Gaussian distributed constellation remains unchanged under arbitrary phase shifts, hence optimizing the power of data streams is sufficient when Gaussian input is assumed. However, a finite constellation generally changes under a phase shift. After superposition, the sum constellation observed by each user depends on the phase shifts applied to the component signals. Different sum constellations lead to different performance. Hence initially both the power and the phase shift applied to each stream need be optimized in order to maximize the performance under finite constellations. Fig. \ref{search_space_SIC} and \ref{search_space_non_SIC} evaluate the CC sum-rate achieved by two searching strategies: 
\begin{enumerate}
    \item To exhaustively search for $t^\star$ and the optimal phase shift between the common and private streams that optimize (\ref{CC_sum_rate_SIC}) or (\ref{CC_sum_rate_no_SIC}), depending on the availability of SIC.
    \item To search among only a small number of samples in $[0,1]$ for one that maximizes (\ref{CC_sum_rate_approx_SIC}) or (\ref{CC_sum_rate_approx_no_SIC}), depending on the availability of SIC, without optimizing the phase shifts applied to the streams.
\end{enumerate}
It can be observed that searching $t^\star$ among $\{0, 0.2, 0.4, ...,1\}$ without any control on the phase shifts leads to a near-optimal performance. We conclude that, given enough freedom on designing $t$, optimizing the phase provides very little performance improvement and hence we neglect the phase shifts in the proposed design. Also, taking (\ref{CC_sum_rate_approx_SIC}) and (\ref{CC_sum_rate_approx_no_SIC}) as objective functions, instead of (\ref{CC_sum_rate_SIC}) and (\ref{CC_sum_rate_no_SIC}), significantly reduces the computational complexity of determining $t^\star$, since there is no need to compute the expectations in (\ref{CC_sum_rate_SIC}) and (\ref{CC_sum_rate_no_SIC}), but poses very little performance loss. As a consequence, the second strategy with searching space $\{0, 0.2, 0.4, ...,1\}$ was applied to generate numerical results in Section. \ref{results}.
\end{remark}

\section{Transceiver designs for RSMA} 
In this section, transceiver designs are proposed for RSMA based on a BICM approach. 

\subsection{Transmitter Design}
Fig. \ref{Tx} depicts the proposed RSMA transmitter architecture, where FEC, $\Pi$ and $M$ represent forward error correction coding, interleaving and mapping (modulation) correspondingly. Message splitting and combining can be achieved by de-multiplexing and multiplexing the information bits. The common message, which is generated by multiplexing $W_{\text{c},1}, W_{\text{c},2},..., W_{\text{c},K}$, and private messages, $W_{\text{p},1}, W_{\text{p},2},..., W_{\text{p},K}$, are encoded, interleaved and mapped separately, to generate blocks of common and private stream symbols $\{s_\text{c},s_1,...,s_K\}$. The precoders then mix the common and private streams before transmission.

\subsection{Receiver Designs}
We introduce five RSMA receiver designs to provide more possibilities in practice, e.g. different complexity and performance. Furthermore, the receivers are classified into SIC and non-SIC type receivers and hence are also specific implementations that push practical systems towards the information theoretic limits characterized by ({\ref{CC_sum_rate_SIC}}) and ({\ref{CC_sum_rate_no_SIC}}) respectively.
\subsubsection{Hard CWIC}
\begin{figure*}[!t]
    \begin{minipage}{0.5\textwidth}
      \centering
      \includegraphics[width=0.7\textwidth]{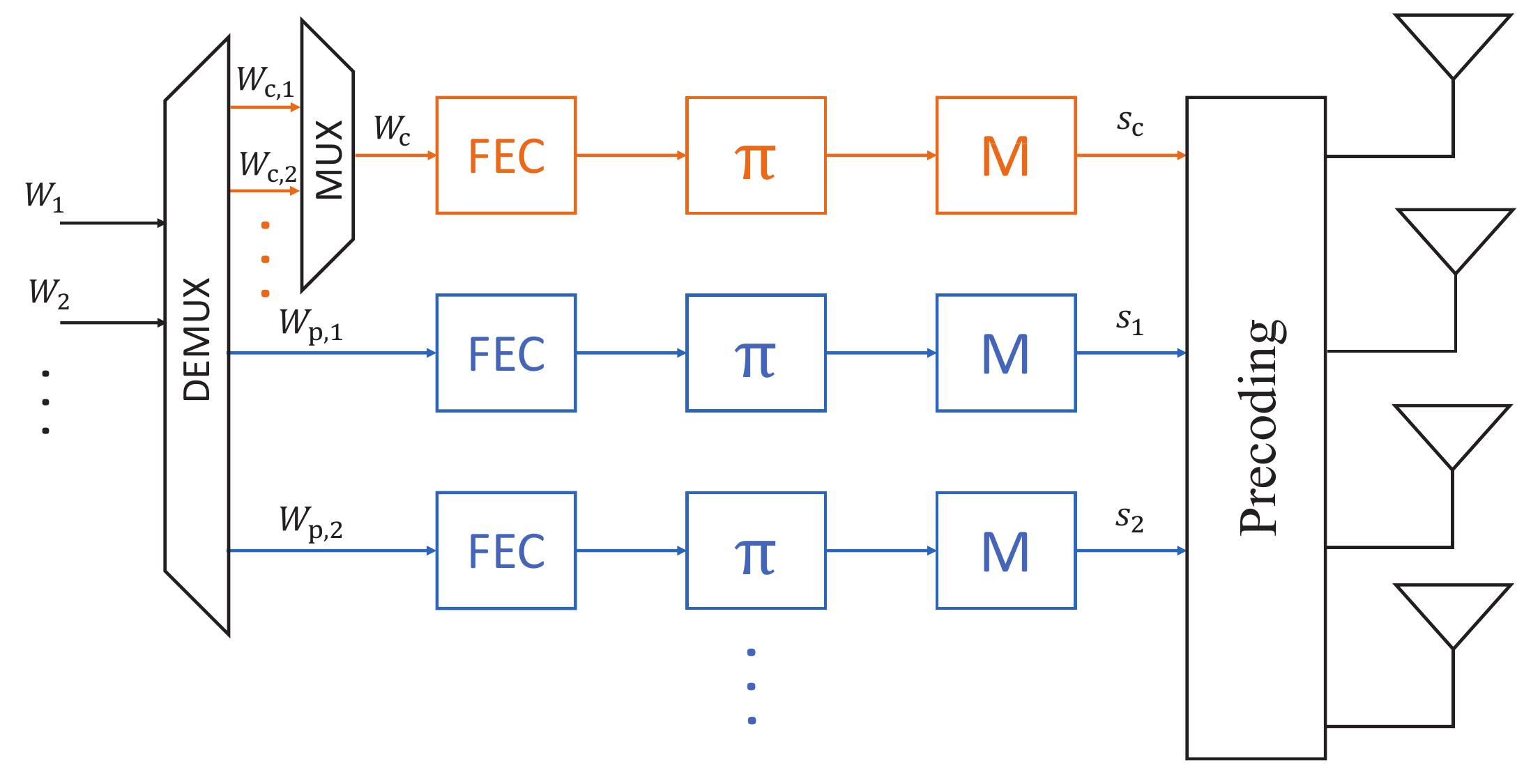}%
      \vspace{-0.4cm}
      \caption{Transmitter architecture.}
      \label{Tx}
      \vspace{0.2cm}
      \end{minipage}
      \begin{minipage}{0.5\textwidth}
      \centering
      \includegraphics[width=0.7\textwidth]{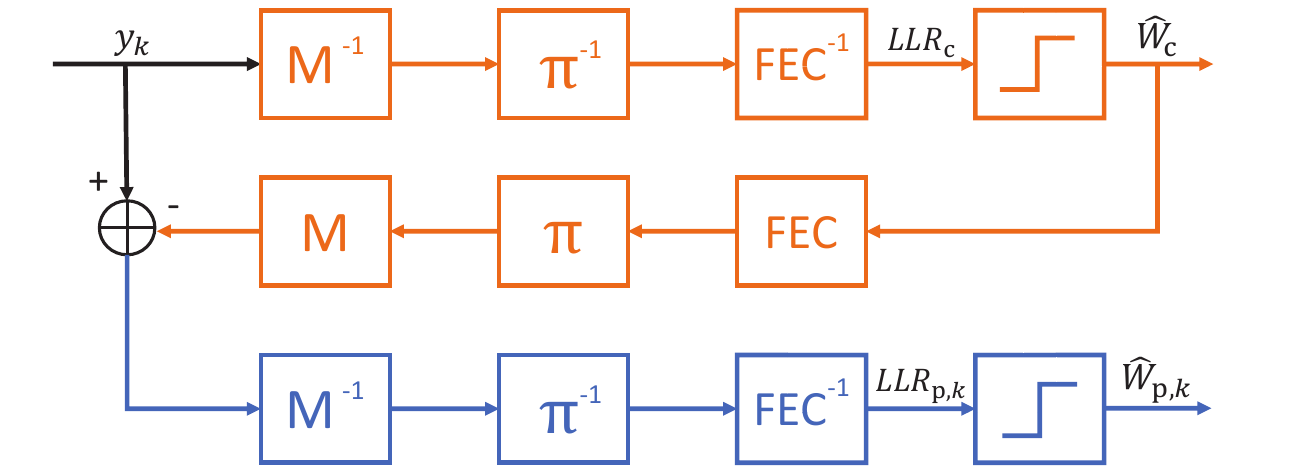}%
      \vspace{0.35cm}
      \caption{Hard CWIC.}
      \label{Hard CWIC}
      \end{minipage}
      \begin{minipage}{0.5\textwidth}
      \centering
      \includegraphics[width=0.7\textwidth]{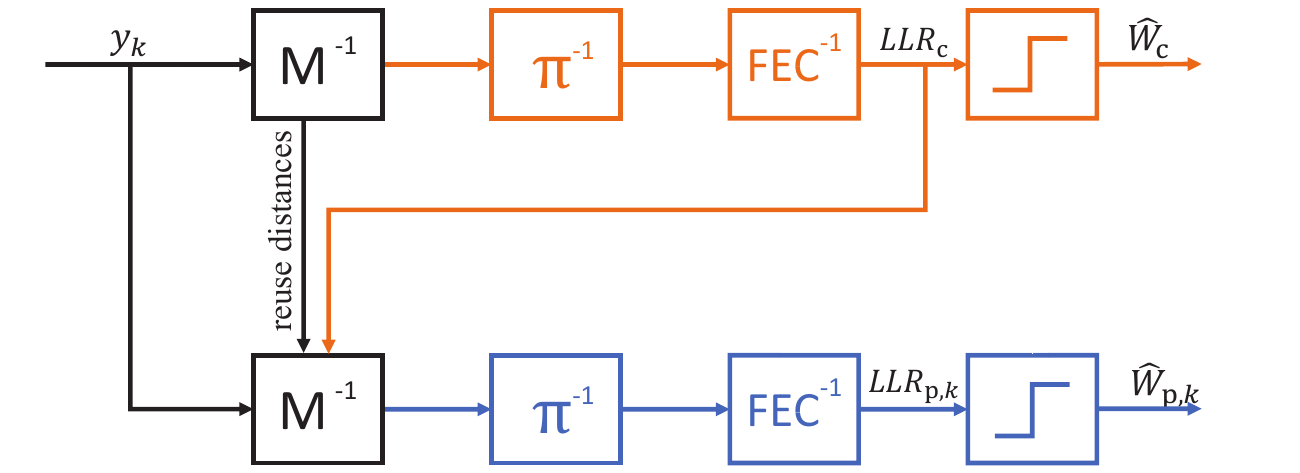}%
      \vspace{0.35cm}
      \caption{Soft CWIC 1.}
      \label{Soft CWIC 1}
      \end{minipage}
      \begin{minipage}{0.5\textwidth}
      \centering
      \includegraphics[width=0.7\textwidth]{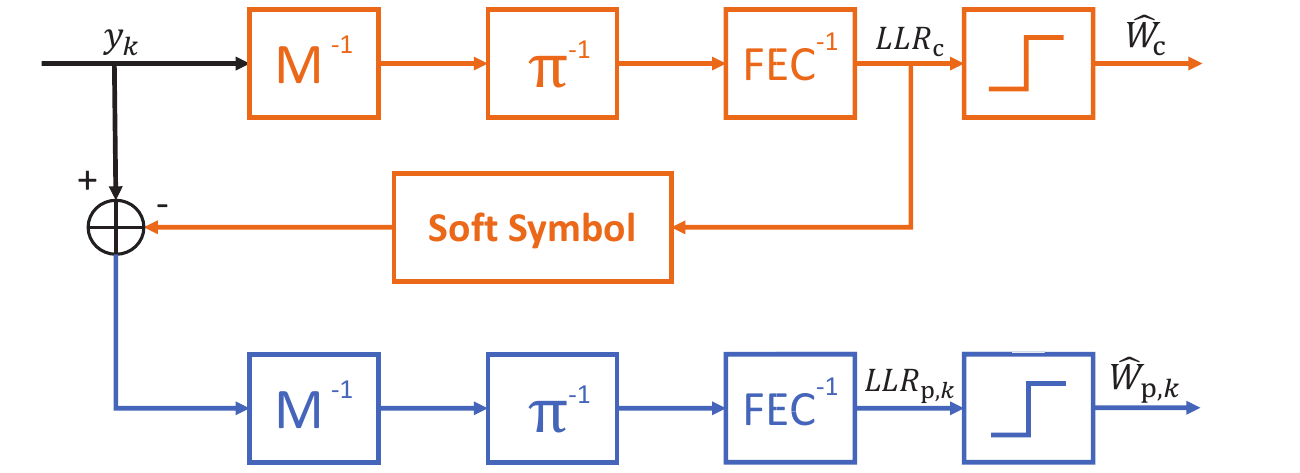}%
      \caption{Soft CWIC 2.}
      \label{Soft CWIC 2}
      \end{minipage}
      \begin{minipage}{0.5\textwidth}
      \centering
      \includegraphics[width=0.7\textwidth]{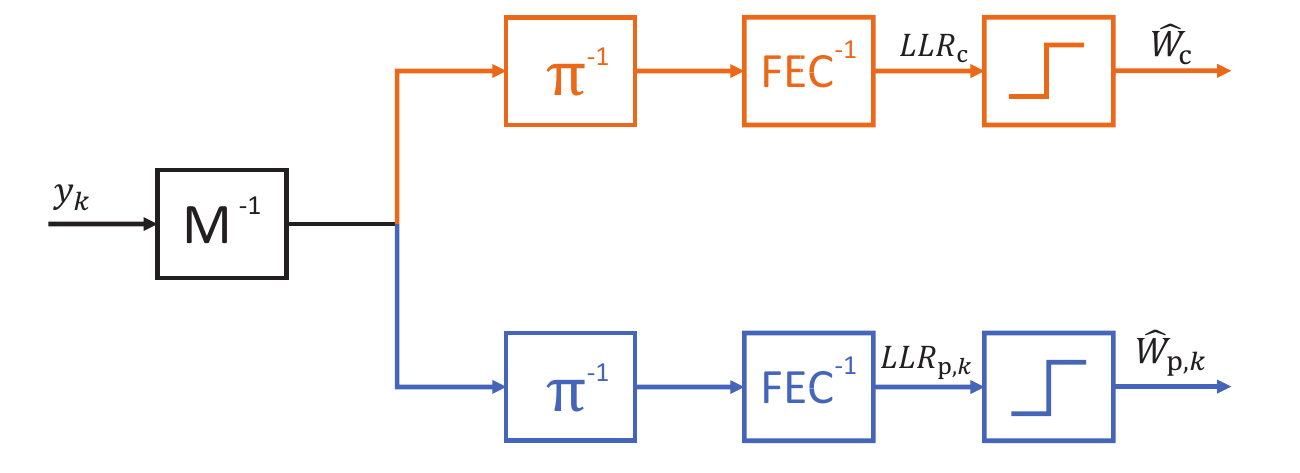}%
      \caption{Joint de-mapper.}
      \label{Joint demapper}
      \end{minipage}
      \begin{minipage}{0.5\textwidth}
      \centering
      \includegraphics[width=0.7\textwidth]{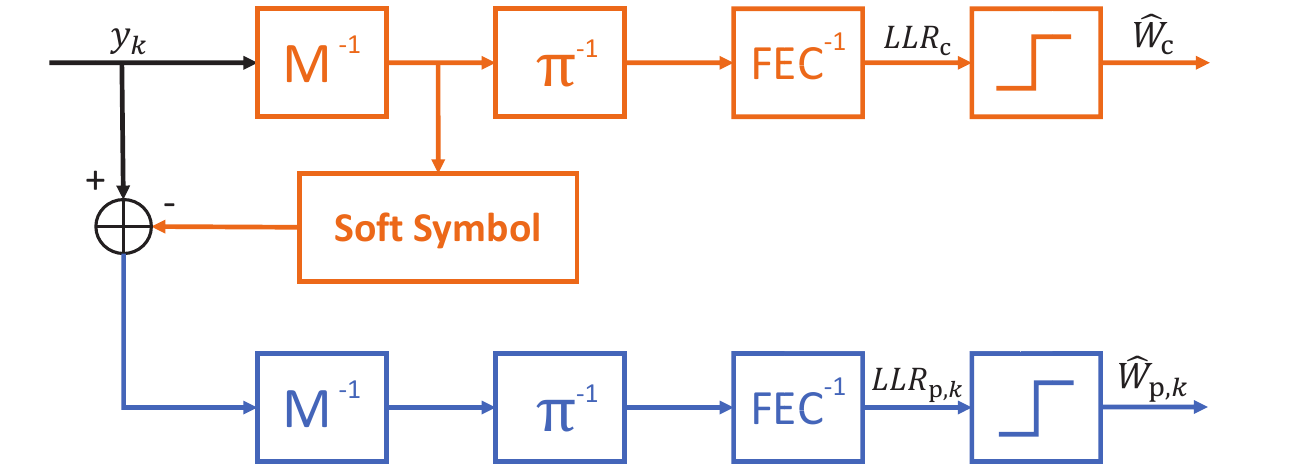}%
      \caption{Soft SLIC.}
      \label{Soft SLIC}
      \end{minipage}
\end{figure*}

Based on the definition of SIC in information theory, hard Code Word level SIC (CWIC) \cite{NAICS,MUST} is the most basic way of implementing SIC in practice as well as the most common RSMA receiver in iterature\mbox{\cite{Dizdar,Dizdar_2,Yin,Chen_and_Mi}}. The structure of a hard CWIC is depicted in Fig. \ref{Hard CWIC}, where $M^{-1}$, $\Pi^{-1}$ and $\text{FEC}^{-1}$ represent de-mapper, de-interleaver and decoder respectively\footnote{In the receiver diagrams, $M^{-1}$ in orange or blue represents Gaussian de-mapper for the common or private stream respectively, i.e. it marginally de-maps one stream by considering the rest of the signals as Gaussian noise. $M^{-1}$ in black represents joint de-mapper which jointly de-maps all the received signals and obtains soft-estimates for the codebits carried by either the common stream or the private stream or both.}. The receiver first de-maps the common stream symbols from $y_k$ by treating the private symbols as Gaussian noise and obtains the Log-Likelihood Ratios (LLRs) of the coded common stream bits. Next, the LLRs are de-interleaved and decoded to obtain the LLRs of the uncoded common stream bits. Then, hard decisions are applied to the LLRs to obtain estimates on $W_\text{c}$, $\Tilde{W}_\text{c}$, followed by re-encoding, re-interleaving and re-mapping $\Tilde{W}_\text{c}$ to obtain estimates on a block of common stream symbols, $\Tilde{s}_\text{c}$. Then, a block of $\mathbf{h}^H\mathbf{p}_\text{c}\Tilde{s}_\text{c}$ is subtracted from the received signal block to cancel the effect of common stream. Finally, the remaining signals, $y_k-\mathbf{h}^H\mathbf{p}_\text{c}\Tilde{s}_\text{c}$, pass through de-mapping, de-interleaving and decoding to obtain the uncoded bits carried by the private stream.

Hard CWIC has a known issue on error propagation, i.e. if the decoding of the common stream is unsuccessful, errors exist in the cancellation process and severely harm the decoding for the private stream. To mitigate this issue, two extended CWIC designs exploiting, instead of the hard information, the soft information from the common stream decoding stage to perform cancellation, are introduced as follows.

\subsubsection{Soft CWIC 1}
The principle of soft CWIC 1 is depicted in Fig. \ref{Soft CWIC 1}. The receiver first jointly de-maps the common stream and the desired private stream but only computes the common stream coded-bit-LLRs\footnote{Here joint de-mapping with marginal outputs is used instead of marginal de-mapping for two reasons: 1) Joint de-mapping is the optimum way of detecting signals in presence of interference while marginal de-mapping treats the discretely distributed interference as Gaussian noise and hence is sub-optimal; 2) If the prior knowledge on the interference distribution is applied in the LLR computation, using joint de-mapping does not introduce more complexity\cite{DVB}.}. As with hard CWIC, the LLRs are de-interleaved and decoded to obtain uncoded-bit-LLRs of the common stream. Assuming the decoder also updates the coded-bit-LLRs\footnote{This is true for many decoder designs used in practice, e.g. belief propagation decoders.}, the probability distribution of a block of common stream symbols, $P(s_\text{c})$, is then computed. Note that this estimate on $P(s_\text{c})$ is more accurate than the one obtained from the de-mapper, as knowledge of the error correction coding is utilized. To de-map the private stream symbols, the receiver performs another joint de-mapping but also uses $P(s_\text{c})$ as prior information. The computation results on distances from the previous de-mapping can be reused at this stage, hence the computational complexity of the second de-mapping is negligible. Following the de-mapping, the private stream is then de-interleaved and decoded. Readers are referred to \cite{DVB} for more details.

\subsubsection{Soft CWIC 2}
With soft CWIC 1, joint de-mapping for the common and private symbols potentially poses high computational complexity. To reduce this computational burden, soft CWIC 2 is introduced and depicted in Fig. \ref{Soft CWIC 2} \cite{Soft_SIC}. In this design, the receiver first de-maps the common stream by treating the private stream symbols as Gaussian noise. $P(s_\text{c})$ is obtained after de-interleaving and decoding in a similar way as with soft CWIC 1. Unlike soft CWIC 1, $P(s_\text{c})$ is used to compute the mean (also called the soft symbol estimates), $\Tilde{s}_\text{c}$, and the variance, $\sigma_{s_\text{c}}^2$, of the common symbols. Then de-mapping for the private symbols are performed by treating the common stream symbols plus noise as an effective noise following the distribution $\mathcal{CN}(\mathbf{h}_k^H\mathbf{p}_\text{c}\Tilde{s}_\text{c}, \sigma^2 + \sigma_{s_\text{c}}^2)$. To do so, the receiver removes $\mathbf{h}_k^H\mathbf{p}_\text{c}\Tilde{s}_\text{c}$ from $y_k$ then calculates the LLRs by taking $\sigma^2 + \sigma_{s_\text{c}}^2$ as the noise variance. In comparison to soft CWIC 1, soft CWIC 2 avoids any joint de-mapping and utilizes $P(s_\text{c})$ in a simpler way and hence is more computationally efficient. 

\subsubsection{Joint De-mapper}
Fig. \ref{Joint demapper} depicts a joint de-mapper \cite{DVB} based receiver design for RSMA. The receiver first jointly de-maps signals from common and private streams. The demodulation outputs are then de-multiplexed into coded-bit-LLRs of common and private streams. Next, the LLRs are separately interleaved and fed into decoders to obtain uncoded bits of the common and private streams.

\subsubsection{Soft SLIC}
To avoid the potentially high complexity brought by the joint de-mapper, a soft Symbol Level Interference Cancellation (SLIC) design \cite{BICM} is introduced as a substitute, as depicted in Fig. \ref{Soft SLIC}. The process is similar to the soft CWIC 2 design mentioned above. The only difference is that the prior knowledge of the common stream symbols is not from the decoder, but the de-mapper for the common stream. Another way to interpret soft SLIC is to view several steps, $M^{-1} \rightarrow \text{Soft Symbol} \rightarrow \text{Cancellation} \rightarrow M^{-1}$, as a simplified joint de-mapping process. Indeed, it de-maps the received signals into soft-estimates on the common and private codebits. Instead of jointly de-mapping the two streams, two marginal de-mapping blocks lead to lower computational burden.



\subsection{Comments on different receivers}

\subsubsection{Complexity}
We summarize the complexity of different receiver designs under three criteria, namely computational complexity, additional buffer size and additional delay. We recognize that the joint de-mapper leads to the simplest structure and the shortest delay among others and hence measure the buffer size and delay in terms of the extra requirement in comparison with joint de-mapper. The three criteria are hence explained as follows:
\begin{itemize}
    \item Computational complexity: The majority of the computation burden comes from de-mapping and decoding. For comparative purposes, we focus on the de-mapping complexity and neglect computation burden from decoding as decoding process of all the receivers is identical, i.e. all the receivers decode the common stream and the desired private streams.
    \item Additional buffer size: This refers to extra buffers required to perform cancellation in comparison to joint de-mapper. For CWIC receivers, the decoded common stream message need to be stored before performing cancellation. The required buffer size depends on the number of coded bits during one transmission and the type of information to be stored, i.e. hard or soft.
    \item Additional delay: This refers to extra time required to perform cancellation in comparison to joint de-mapper. Apart from all the process of joint de-mapper, CWIC receivers need extra time to perform re-interleaving and re-modulation (or more generally, converting bit level information into symbol level information) before cancellation. Note that the time consumption on joint de-mapping and soft SLIC can be shortened by parallelized decoding of common and private message, but this is not valid for CWIC receivers due to the existence of cancellation process.
\end{itemize}
Let $\mathcal{X_\text{c}}$ and $\mathcal{X_\text{p}}$ be the constellations for common and private streams, $F$ be the number of bits required for storing soft information, i.e. the LLRs, and $B$ be the blocklength in terms of symbols. With a quantitative evaluation of receiver complexity given in Table. \ref{Rx_complexity}, it can be concluded that the soft CWIC 1 and the soft SLIC are of the highest and lowest complexity respectively. 

\begin{table*}
\caption{Classification and complexity of different receiver designs.}
\centering
\begin{tabular}{|c|c|c c c|} 
 \hline
  & Receiver & Computational complexity & Additional buffer size & Additional delay \\ [0.5ex] 
 \hline\hline
 \multirow{3}{6em}{\;\;\;\;SIC Type} & Hard CWIC & $\mathcal{O}\Big(|\mathcal{X_\text{c}}| + |\mathcal{X_\text{p}}|\Big)$ & $B \log_2 |\mathcal{X_\text{c}}|$ & Interleaving \& Mapping \\ 
 \cline{2-5}
 & Soft CWIC 1 & $\mathcal{O}\Big(|\mathcal{X_\text{c}} \times \mathcal{X_\text{p}}|\Big)$ & $F B \log_2 |\mathcal{X_\text{c}}|$ & Interleaving \& Mapping   \\
 \cline{2-5}
 & Soft CWIC 2 & $\mathcal{O}\Big(|\mathcal{X_\text{c}}| + |\mathcal{X_\text{p}}|\Big)$ & $F B \log_2 |\mathcal{X_\text{c}}|$ & Interleaving \& Mapping   \\
 \hline
 \multirow{2}{6em}{Non-SIC Type}
 & Joint De-mapper & $\mathcal{O}\Big(|\mathcal{X_\text{c}} \times \mathcal{X_\text{p}}|\Big)$ & 0 & -\\ 
 \cline{2-5}
 & Soft SLIC & $\mathcal{O}\Big(|\mathcal{X_\text{c}}| + |\mathcal{X_\text{p}}|\Big)$ & 0 & -\\
 \hline
\end{tabular}
\label{Rx_complexity}
\end{table*}

\subsubsection{SIC and non-SIC receivers}
The above-mentioned receiver designs can be classified into two groups according to their underlying principle. In particular, hard CWIC, soft CWIC 1 and soft CWIC 2 utilize the common stream decoder outputs in the process of decodng the private stream and hence are classified as SIC-type receivers. Joint de-mapping and soft SLIC decode the private stream without using the output of the common stream decoder by treating the common stream as noise. Hence they are classified as non-SIC-type receivers. This classification is also illustrated in Table. {\ref{Rx_complexity}}.

Jointly de-mapping the desired signal and interference is equivalent to treating interference as noise. Two supporting evidences are given:
\begin{enumerate}
    \item From an information theoretic perspective, approaches to handling interference can be classified into: 1) decoding interference and 2) treating interference as noise. From Fig. \ref{Joint demapper}, it is easy to see that joint de-mapping does not perform SIC or joint decoding, hence it does not belong to the category of decoding interference.
    \item From a detection problem perspective, detecting discrete signal in the presence of discrete interference and Gaussian noise is equivalent to jointly de-mapping the desired signal and interference. A proof is given in Appendix B.
\end{enumerate}

A common confusion exists between two ideas: 1) treating interference as noise and 2) treating interference as Gaussian noise. With finite constellations, interference is not Gaussian, and hence the sum of interference and noise is not Gaussian. The optimal way of treating interference as noise is to use its exact probability distribution when de-mapping as shown in Appendix B. This leads to joint de-mapping. However, in practice, it is computationally more efficient to pretend that the interference is Gaussian, and hence the sum of interference and noise is also Gaussian. This leads to receivers treating the sum of interference and noise as as an effective Gaussian noise (such as receivers for downlink MU-MIMO), which could result in a huge performance loss in comparison to joint de-mapping because the wrong probability distribution for the interference is used. For RSMA, jointly de-mapping the common and the private streams is equivalent to treating the common stream as noise when decoding the private stream.

In light of the classification of SIC type and non-SIC type receivers, it is evident that, depending on the type of the receivers in use, the system should be designed to maximize the corresponding CC sum-rate (with or without SIC) in order to optimize the performance.

\section{Numerical results}\label{results}
We evaluate the proposed designs through numerical simulations. In what follows, the channel model in use is first introduced, followed by achievable rate evaluation and finally LLS evaluation of the precoder and receiver designs. Throughout this section, we assume $\theta=\pi/3$, $N_t=4$ for two-user cases and $N_t=6$ for three-user cases.

\begin{figure}[!h]
      \centering
      \includegraphics[width=2.6in]{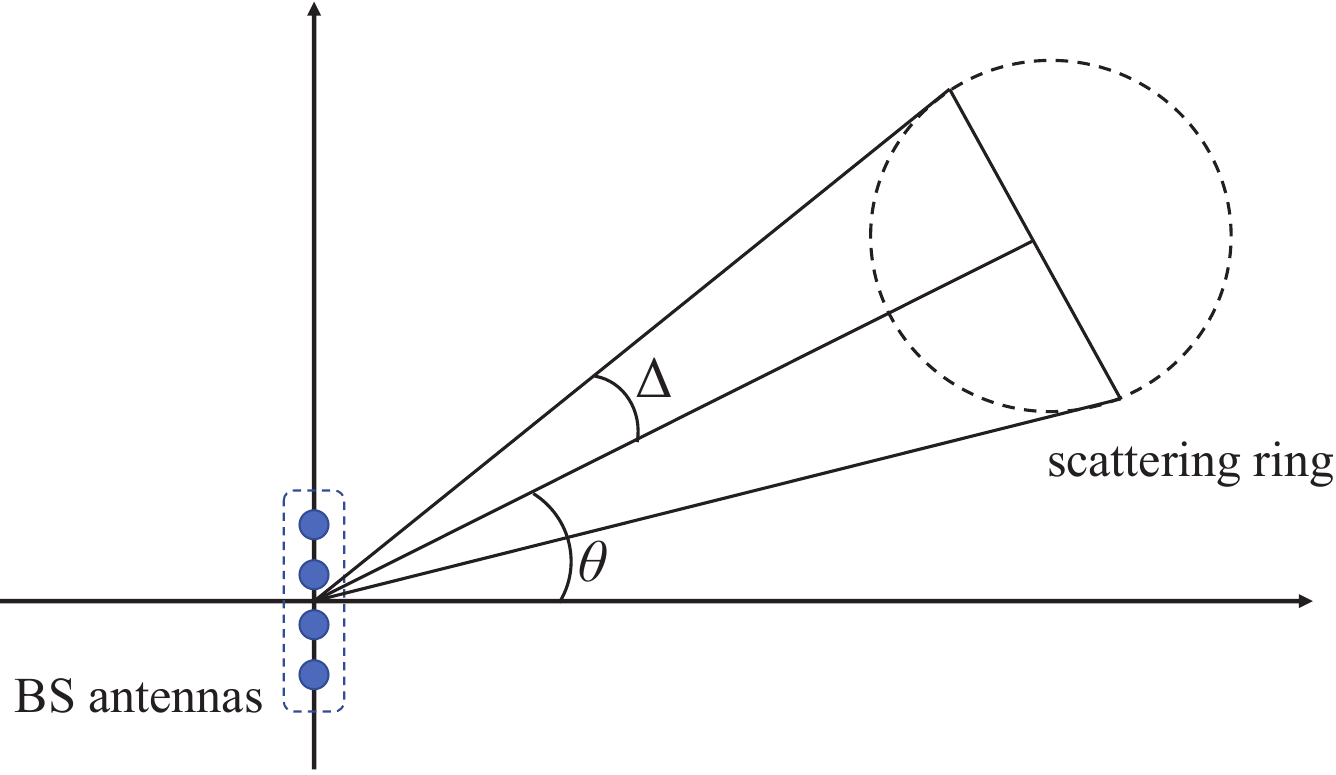}
      \caption{A Base Station (BS) antenna array and a scattering ring centered at AoD $\theta$ with two sided angular spread $\Delta$.}
      \label{One-ring}
\end{figure}

\begin{figure*}[!h]
    \centering
    \subfigure[$K=2$, $\Delta=\pi/18$, QPSK for all the streams.]{
    \includegraphics[width = 0.295\textwidth]{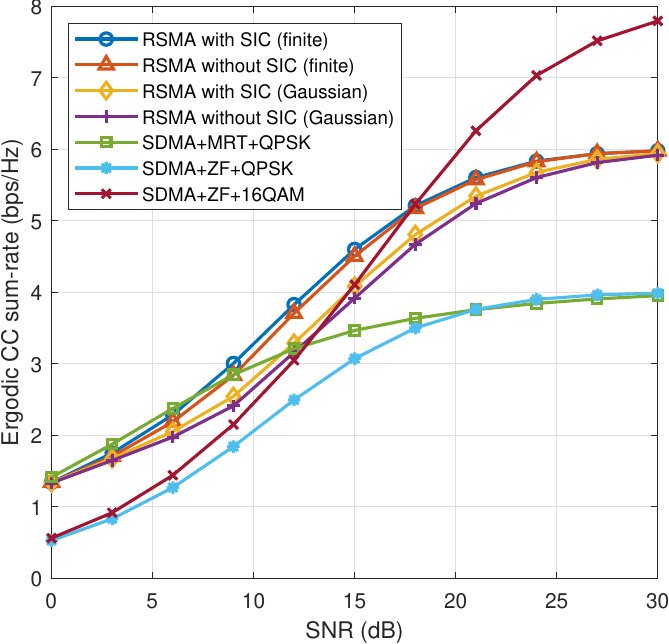}}\hspace{0.01\textwidth}
    \hspace{0.2cm}
    \subfigure[$K=2$, $\Delta=\pi/18$, 16QAM for all the streams.]{
    \includegraphics[width = 0.3\textwidth]{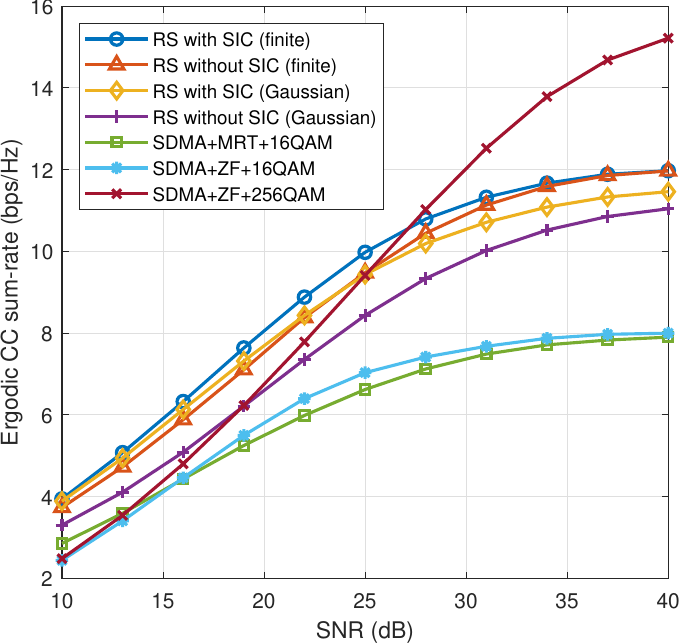}}\hspace{0.01\textwidth}
    \hspace{0.2cm}
    \subfigure[$K=3$, $\Delta=\pi/9$, 16QAM for common streams, QPSK for private streams.]{
    \includegraphics[width = 0.3\textwidth]{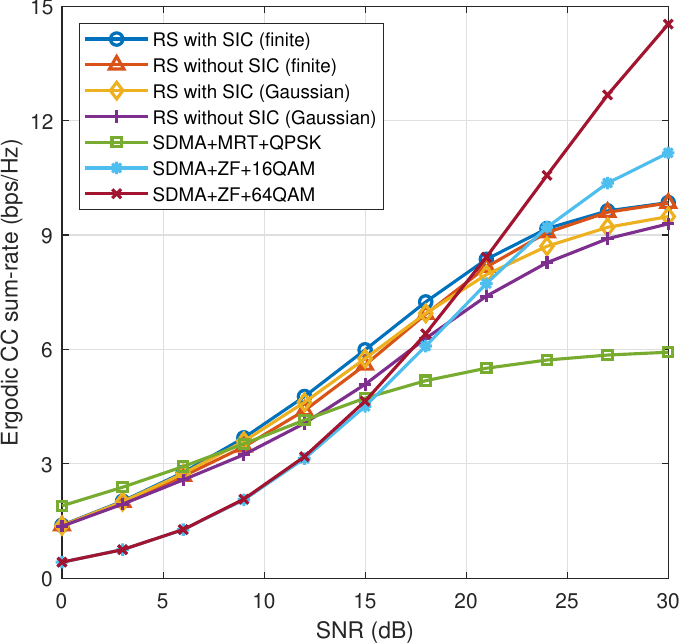}}
    \caption{CC sum-rate comparison for different schemes.}
    \label{CC_sum_rate_comp}
\end{figure*}

\subsection{Channel Model}
The spatially correlated Rayleigh-fading channel is assumed, i.e. $\mathbf{h}_k \sim \mathcal{CN}(\mathbf{0}_{N_t \times 1},\;\mathbf{R}_k)$ and $\mathbf{R}_k \in \mathbb{C}^{N_t \times N_t}$ is the the channel covariance matrix for user-$k$. Applying eigendecomposision to $\mathbf{R}_k$, one can have 
\begin{equation}
    \mathbf{R}_k = \mathbf{U}_k \mathbf{\Lambda}_k \mathbf{U}_k^H, 
\end{equation}
where $\mathbf{\Lambda}_k$ is a $r_k \times r_k$ diagonal matrix of the nonzero eigenvalues of $\mathbf{R}_k$ and $\mathbf{U}_k \in \mathbb{C}^{N_\text{t} \times r_k} $ is a matrix containing the eigenvectors of $\mathbf{R}_k$ corresponding to the nonzero eigenvalues. Using Karhunen-Loeve representation, $\mathbf{h}_k$ can be expressed in the form 
\begin{equation}
    \mathbf{h}_k = \mathbf{U}_k\mathbf{\Lambda}_k^{\frac{1}{2}}\mathbf{w}_k,
\end{equation}
where $\mathbf{w}_k \in \mathbb{C}^{r_k \times 1} \sim \mathcal{CN}(\mathbf{0,\;I})$.

We further consider the one-ring scattering model \cite{channel}, where the underlying assumption is that, between the transmitter and user-$k$, there is no line-of-sight path and the angle-of-departure (AoD) of all the scatters is uniformly distributed around a centre AoD, $\theta_k$, with an angular spread of $\Delta_k$, as depicted in Fig. \ref{One-ring}. Assuming a uniform and linear antenna array centered at the origin and along the $y$-axis with the spacing of antenna elements to be $\lambda/2$, as depicted in Fig. \ref{One-ring}, the $(m,n)$-th elements of $\mathbf{R}_k$ is given by
\begin{equation}
    [\mathbf{R}_k]_{m,n} = \frac{1}{2\Delta_k} \int_{\theta_k-\Delta_k}^{\theta_k+\Delta_k}
    e^{-j\pi (m-n) \sin{\alpha}} d\alpha.
\end{equation}
For simplicity, we assume $\theta_1=\theta_2=...=\theta_K=\theta$ and $\Delta_1=\Delta_2=...=\Delta_K=\Delta$, which implies that users with similar channel statistics are served at the same time. With this setting, we intend to model scenarios with certain user density. A small $\Delta$ indicates a high density demand area.

\subsection{CC Sum-Rate Comparison}
We verify that the proposed low-complexity precoder design that optimizes (\ref{CC_sum_rate_approx_SIC}) and (\ref{CC_sum_rate_approx_no_SIC}) actually leads to higher CC sum-rate than other schemes. In Fig. {\ref{CC_sum_rate_comp}}, we compare the ergodic CC sum-rate of three schemes, namely RSMA with precoder optimizing the sum-rate assuming Gaussian signalling (with "Gaussian" in the legend), RSMA with precoder optimizing the approximation of CC sum-rate (with "finite" in the legend) and SDMA with either ZF or MRT precoding under different scenarios. Constellations used for different schemes are given in the legend and caption of the figure. The constellations were chosen such that different schemes require similar decoding complexity. It can be observed that RSMA with precoders optimizing the sum-rate does not essentially optimize the CC sum-rate and often underperforms SDMA. Especially when non-SIC receiver is assumed, RSMA by optimizing the sum-rate always under-performs SDMA. On the other hand, the designs that optimize (\ref{CC_sum_rate_approx_SIC}) and (\ref{CC_sum_rate_approx_no_SIC}) lead to nearly the best performance among others at low and medium SNRs. At very low SNR regime, SDMA with MRT precoding slightly outperforms RSMA. However, this performance is hard to achieve in practice. The reason is that CC sum-rate provides the performance upper-bound and, to achieve this, the receiver should at least jointly de-maps the desired signal and interference rather than simply treating interference as Gaussian noise, as discussed in Section IV C. To do so, advanced receivers and extra knowledge on interference is necessary. At high SNR regime, SDMA with high-order constellations outperforms RSMA because the latter saturates at a lower rate. For example, RSMA with QPSK used for both common and private streams transmits at most 6 bits/symbol, while SDMA with two 16QAM streams transmits at most 8 bits/symbol.

The above results lead to two important implications. First, the potential of RSMA can be better discovered when the finite constellations are considered during the system design process. Second, the superiority of RSMA over SDMA remains even under constraints on similar decoding complexity.

\subsection{Power allocation behavior}

\begin{figure}[!h]
    \centering
    \subfigure[Channel 1.]{
    \includegraphics[width = 0.225\textwidth]{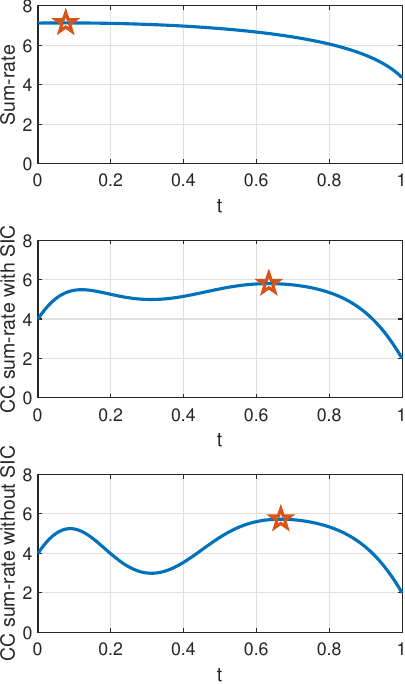}}%
    \hspace{0.2cm}
    \subfigure[Channel 2.]{
    \includegraphics[width = 0.225\textwidth]{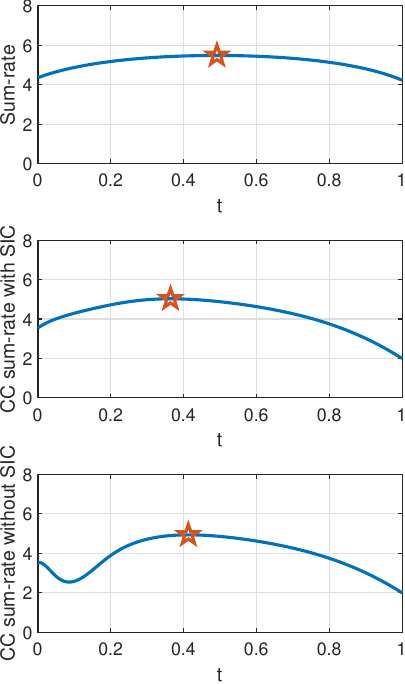}}
    \caption{Sum-rate/CC sum-rate as functions of $t$ with $t^\star$ labeled by $\bigstar$, $K=2$.}
    \label{sum_rate_vs_t}
\end{figure}

\begin{figure*}[!h]
    \centering
    \subfigure[SNR=5dB, QPSK assumed for CC sum-rates.]{
    \includegraphics[width = 0.3\textwidth]{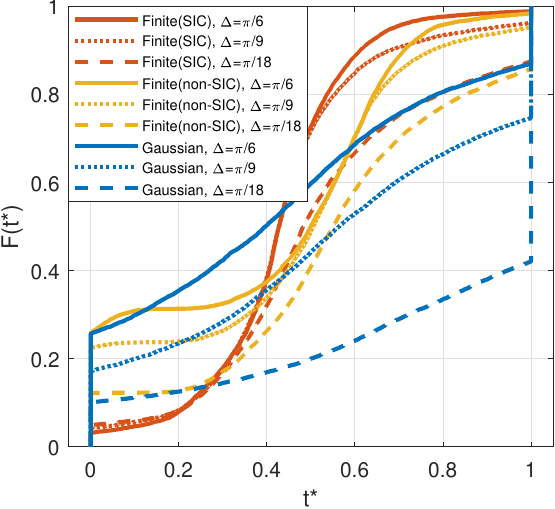}\label{CDF_of_t_5dB_QPSK}}\hspace{0.01\textwidth}
    \hspace{0.2cm}
    \subfigure[SNR=10dB, QPSK assumed for CC sum-rates.]{
    \includegraphics[width = 0.3\textwidth]{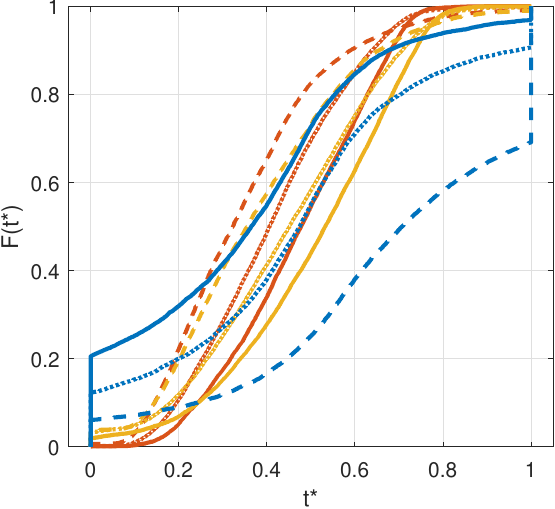}\label{CDF_of_t_10dB_QPSK}}\hspace{0.01\textwidth}
    \hspace{0.2cm}
    \subfigure[SNR=10dB, 16QAM assumed for CC sum-rates.]{
    \includegraphics[width = 0.3\textwidth]{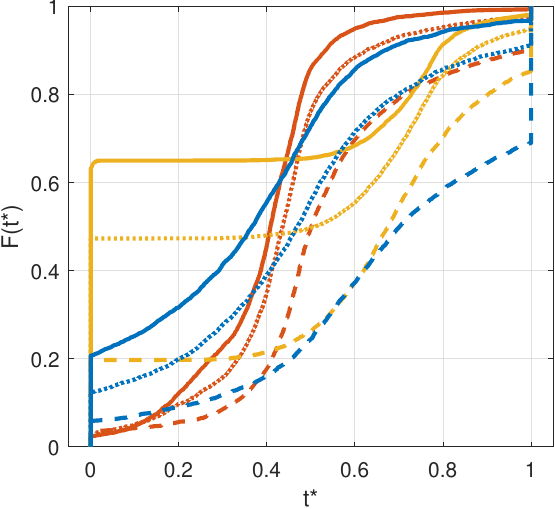}\label{CDF_of_t_10dB_16QAM}}
    \caption{CDF of $t^\star$ under different objectie functions with $K=2$.}
    \label{CDF_of_t}
\end{figure*}

In the following, we show that taking sum-rate or CC sum-rate as the objective function leads to distinct choices on the power allocation factor, $t$. We first consider two channel realizations given in Appendix C. Fig. {\ref{sum_rate_vs_t}} shows, under the two realizations, the sum-rate, approximated CC sum-rate with SIC receivers and approximated CC-sum-rate with non-SIC receivers, i.e. ({\ref{sum_rate}}), ({\ref{CC_sum_rate_approx_SIC}}) and ({\ref{CC_sum_rate_approx_no_SIC}}) respectively, as a function of $t$, where $t^\star$ for ({\ref{CC_sum_rate_approx_SIC}}) and ({\ref{CC_sum_rate_approx_no_SIC}}) are obtained from exhaustive search. It can be observed that $t^\star$ under different objective functions is indeed different. Also, it is can be illustrated that ({\ref{sum_rate}}) is convex over $t$ while ({\ref{CC_sum_rate_approx_SIC}}) and ({\ref{CC_sum_rate_approx_no_SIC}}) are not. 

Fig. {\ref{CDF_of_t}} depicts the CDF of $t^\star$, denoted by $F(t^\star)$, under different objective functions and channel conditions. 
It shows that the statistics of $t^\star$ can be very different under different objective functions. One interesting observation can be made from the variation of CDF with different $\Delta$. As $\Delta$ decreases, the system with Gaussian signalling always allocates more power to the common stream as this leads to less interference. However, with finite constellations, this is not always the case. For example, in Fig. {\ref{CDF_of_t_10dB_QPSK}}, the system allocates less power with a decreasing $\Delta$. This is because the rate of finite constellations is upper-bounded and saturates after certain SNR. With smaller $\Delta$, the system can achieve the saturating point for the common stream with a smaller amount of power and it is more beneficial to allocate the rest of the power to the private streams as they have not saturated yet. In Fig. {\ref{CDF_of_t_10dB_16QAM}}, the system again allocates more power to the common streams as $\Delta$ drops because 16QAM requires more power than QPSK to saturate in CC rates.

\subsection{Link-Level Simulation}
Link-level simulation was performed to provide a more comprehensive evaluation of different precoders and receivers. In the following, fast fading and block fading channels\footnote{Fast fading refers to i.i.d. channel realizations for each symbol. Block fading refers to channel realizations that are i.i.d. from block to block and identical within the same block.} with perfect CSIT are considered. For error correction code, polar code was chosen for its good error-correcting capability. Belief propagation decoding was used as the decoding technique since it is widely used in practice. All the results were obtained by averaging the Bit Error Rate (BER) performance of 50000 blocks. Symbol block length is 512 for fast fading channels and 4096 for block fading channels. When block fading is considered, it is assumed that channel is coherent for 32 symbols. Max-log approximation was used for de-mapping as it is common in real applications. The target BER was $10^{-4}$.

\begin{figure}[h!]
\centering
 \subfigure[MCS: QPSK/0.54 for common stream and QPSK/0.3 for private streams.]{
      \includegraphics[width=3.35in]{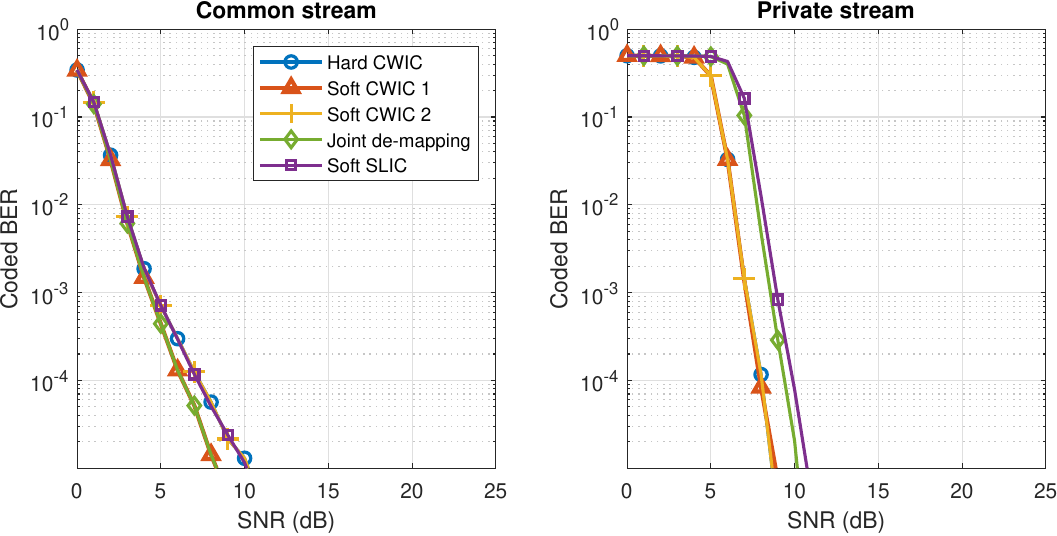}\label{error_prop_good}}
\centering
\vspace{0.3cm}
 \subfigure[MCS: QPSK/0.6 (hard CWIC, soft CWIC 2 and soft SLIC) or QPSK/0.74 (soft CWIC 1 and joint de-mapping) for common stream and QPSK/0.3 for private streams.]{
      \includegraphics[width=3.35in]{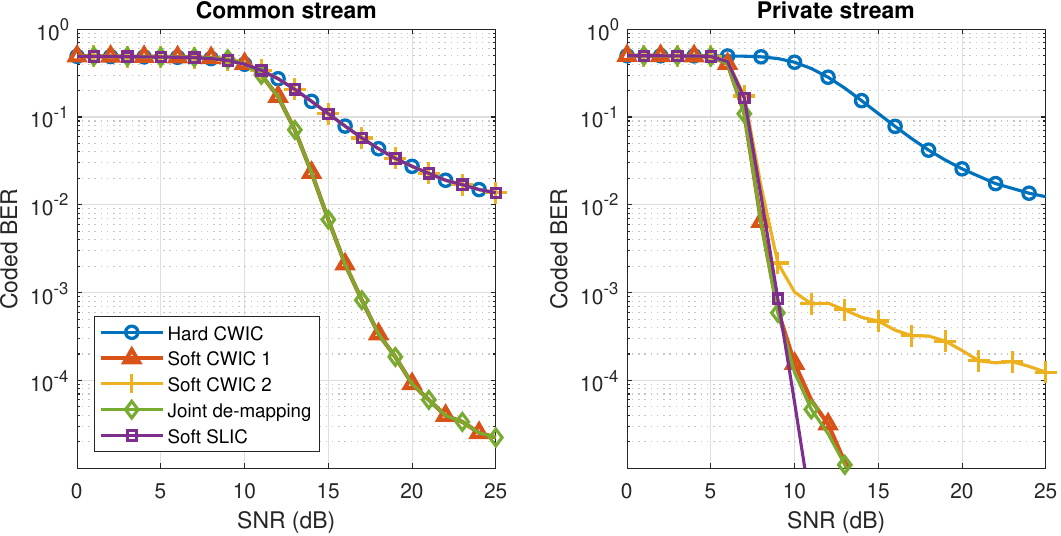}\label{error_prop}}
      \vspace{-0.2cm}
\caption{BER performance of different RSMA receivers with $K=2$ and $\Delta=\pi/9$.}
\end{figure}

We first make a comparison of different receiver designs mentioned in Section IV based on Fig. \ref{error_prop_good} and Fig. \ref{error_prop}. Here, the precoders were designed to maximize the sum-rate. Details on the modulation and coding schemes (MCS) in use are shown in the caption of the figures in the format of "modulation/code rate". We first set the common stream code rate to a medium value such that common and private streams reach $10^{-4}$ BER with similar SNR. This simulates cases where a good operating point is found. In Fig. \ref{error_prop_good}, it can be conclude that the performance of different receivers follow the order: soft CWIC 1 $>$ soft CWIC 2 $>$ hard CWIC $>$ joint de-mapping $>$ soft SLIC. 

\begin{figure*}[h!]
    \centering
    \subfigure[Fast fading channel with $K=2$, $\Delta=\pi/18$ and throughput = 3 bits/symbol. Modulation schemes for RSMA: QPSK for both common and private streams.]{\includegraphics[width=3.35in]{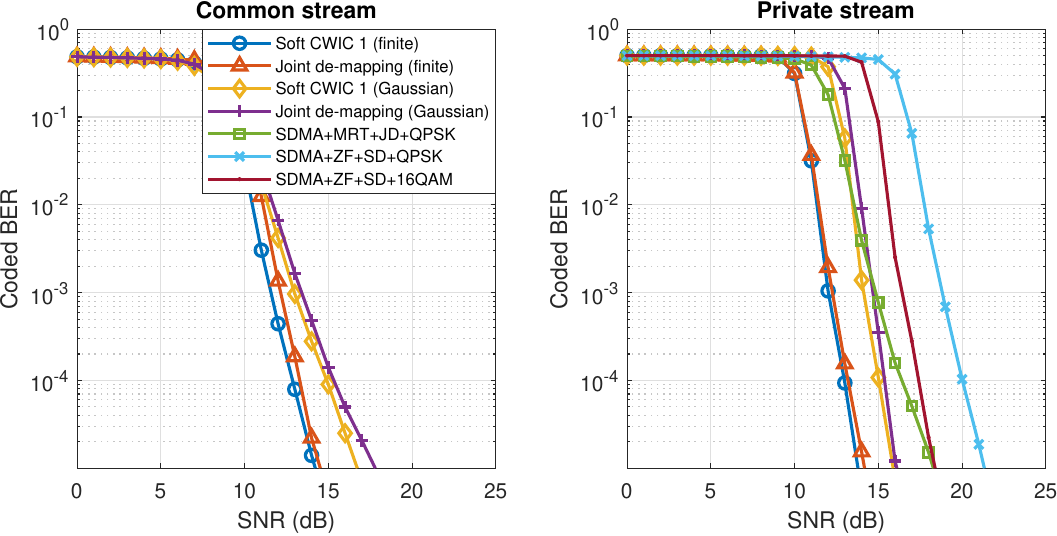}\label{BER_10}}\hspace{0.4cm}
    \centering
    \subfigure[Fast fading channel with $K=2$, $\Delta=\pi/18$ and throughput = 6 bits/symbol. Modulation schemes for RSMA: 16QAM for both common and  private streams.]{\includegraphics[width=3.35in]{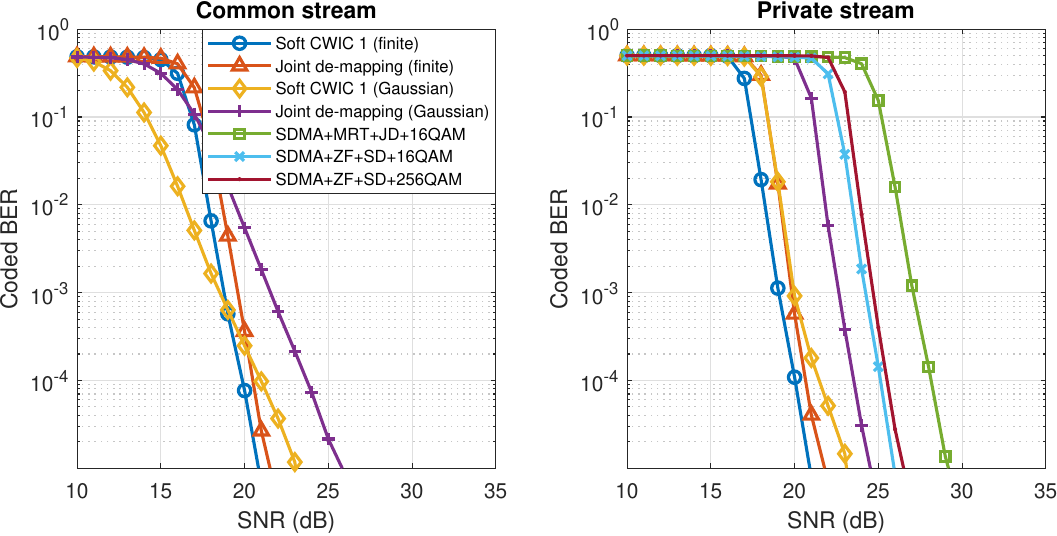}\label{BER_10_16QAM}}
    \centering
    \subfigure[Block fading channel with $K=2$, $\Delta=\pi/18$ and throughput = 3.6 bits/symbol. Modulation schemes for RSMA: QPSK for both common and private streams.]{
      \includegraphics[width=3.35in]{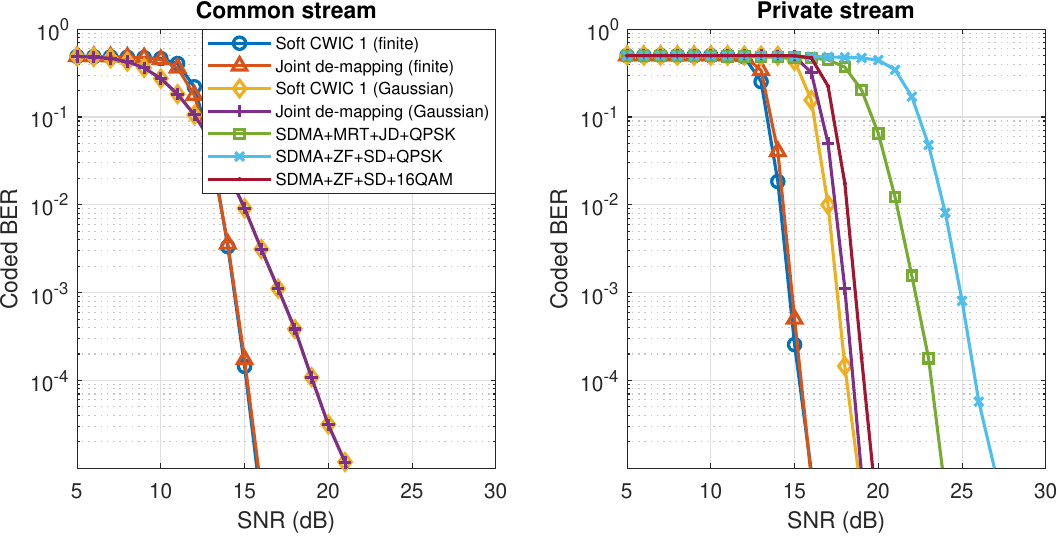}\label{BER_block_fading}}\hspace{0.4cm}
    \centering
    \subfigure[Block fading channel with $K=3$, $\Delta=\pi/9$ and throughput = 5.52 bits/symbol. Modulation schemes for RSMA: 16QAM for common stream and QPSK for private streams.]{\includegraphics[width=3.35in]{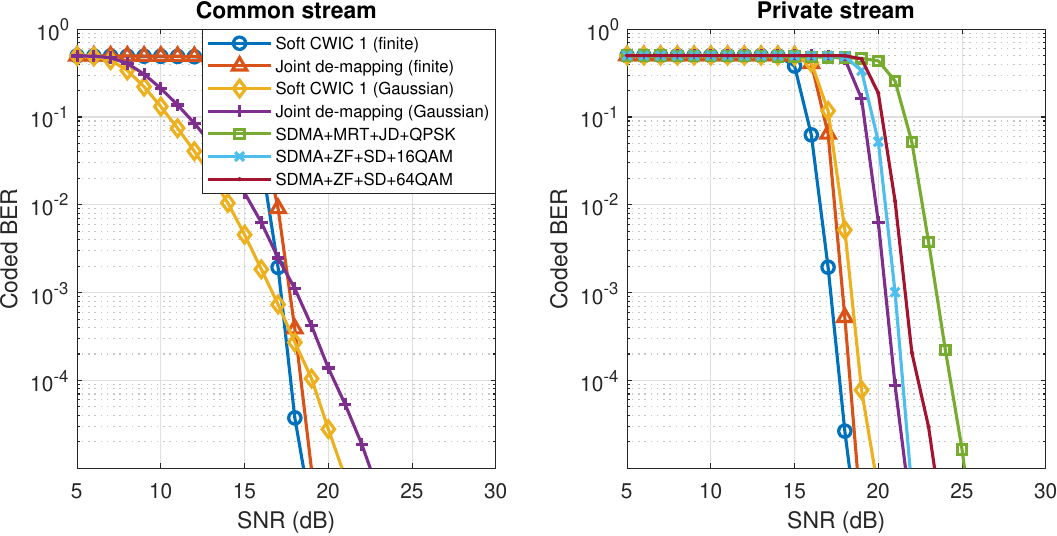}\label{BER_three_user}}
    \caption{BER performance of different schemes.}
    \label{BER_performance}
    \vspace{-0.5cm}
\end{figure*}

We now consider cases where the common stream rate is not properly set and hence errors may exist when decoding the common stream. In these cases, error propagation effects with different receivers become important. Fig. \ref{error_prop} shows the BER performance with different receiver designs when a high error rate occurs in the common stream. The code rate for the common stream was set to be high such that a high BER (around 0.4) occurs when the SNR is 10 dB. When hard CWIC is applied, it can be observed that a similar BER occurs in the private stream due to the error propagation. In contrast, with other receivers, the BER degradation caused by error propagation is very weak or not visible. 

Next, we compare the performance of sum-rate oriented precoder design and CC sum-rate oriented precoder designs through LLS. For clarity, only soft CWIC 1 and joint de-mapping are considered in the following as representatives of SIC-type receivers and non-SIC-type receivers. RSMA schemes with power allocation methods considering sum-rate or CC sum-rate, as proposed in Section III, are indicated by '(Gaussian)' and '(finite)' in the legend. We also included several SDMA-based schemes as benchmarks:
\begin{itemize}
    \item SDMA+ZF+SD+QPSK/16QAM/64QAM/256QAM: The precoders are designed using ZF with uniform power allocation. The receivers apply single user detector (only de-map the desired signal). Both streams are transmitted in QSPK, 16QAM or 256QAM signals.
    \item SDMA+MRT+JD+QPSK/16QAM: The precoders are designed using Maximum Ratio Transmission (MRT) with uniform power allocation. The receivers jointly de-map the desired stream and interference\footnote{As discussed in Section IV C, joint de-mapping is the optimal detector in the presence of finite-constellation interference.}. This receiver design originates from studies on advanced receivers for MU-MIMO in 3GPP\cite{3GPP_Rx}. Both streams are transmitted in QPSK or 16QAM signals. 
\end{itemize}
For clarity, the code rates in use are given by Appendix D.

The BER performance of RSMA with different precoder and receiver designs, as well as three SDMA benchmarks are depicted in Fig. {\ref{BER_10}}-{\ref{BER_three_user}}, with maximum throughput and modulation schemes in use for RSMA are given in the caption of the figures. In particular, Fig. {\ref{BER_10}} and Fig. {\ref{BER_10_16QAM}} examine scenarios with fast fading and two users. Fig. {\ref{BER_block_fading}} assumes block fading with two users while Fig. {\ref{BER_three_user}} assumes block fading with three users. In each figure, different schemes achieve the same throughput as indicated in the captions. For SDMA schemes, the code rates are uniform across users. For RSMA schemes, the code rates for common and private streams are selected such that both streams achieves the same BER at similar SNRs. It can be observed that RSMA schemes generally outperform SDMA schemes. Among different RSMA schemes, the performance depends on the precoder and receiver designs. In particular, it can be observed that the CC sum-rate oriented RSMA (with "(finite)" in the legends) significantly outperforms the sum-rate oriented RSMA (with "(Gaussian)" in the legends), with either SIC-type or non-SIC-type receivers. For example, in Fig. \ref{BER_10}, for either SIC-type or non-SIC-type receivers, CC sum-rate oriented precoder design performs about 2.5 dB better than sum-rate oriented designs. This again reveals the important fact that the practical capability of RSMA is not well exploited by the sum-rate oriented designs, but better discovered by the CC sum-rate oriented designs.

\begin{figure}[h!]
\centering
      \includegraphics[width=2.7in]{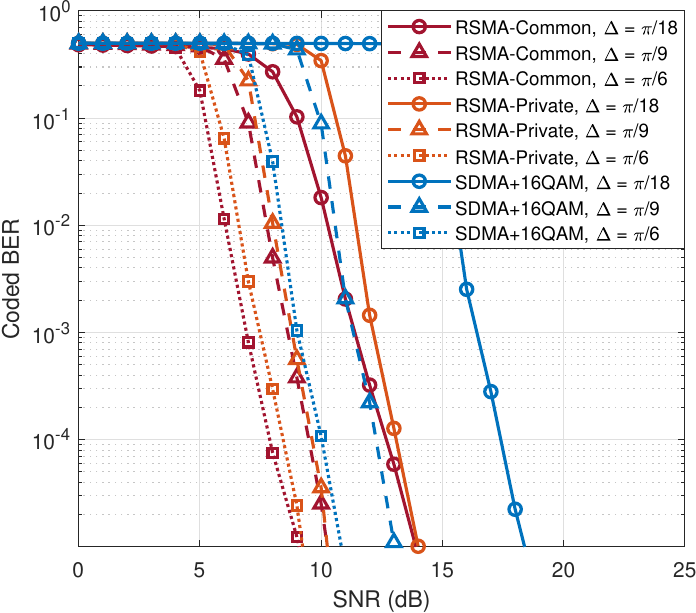}%
      \caption{Comparison between RSMA and SDMA with $K=2$, $\Delta = \pi/18$, $\pi/9$ or $\pi/6$. RSMA uses QPSK for both common and private streams. The throughput of different schemes is 3 bits/symbol.}
      \label{different_delta}
\end{figure}

\begin{figure}[h!]
\centering
      \includegraphics[width=2.6in]{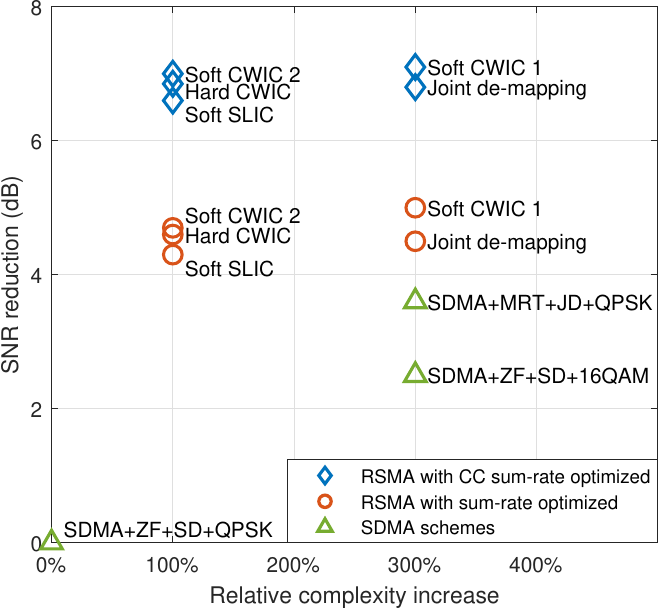}%
      \caption{SNR-complexity trade-off with different schemes under fast fading with $K=2$, $\Delta=\pi/18$ and throughput = 3 bits/symbol. Modulation schemes for RSMA: QPSK for all the streams.}
      \label{SNR_complexity}
\end{figure}

Fig. \ref{different_delta} depicts how the angle spread, $\Delta$, of the scattering ring affects the superiority of RSMA over SDMA. It can be concluded that small angle spreads lead to more benefit being realised from RSMA.

In Fig. \ref{SNR_complexity}, the trade-offs between performance and complexity with different schemes and receivers are depicted based on the LLS results and Table I. SDMA with QPSK constellation was chosen as the reference. SNR reduction was measured in the amount of SNR saved by a scheme when a BER of $10^{-4}$ is achieved. Relative complexity increase was measured in the additional complex distances to be computed per symbol. Without considering the cost on the buffering memory and delay, CC sum-rate oriented RSMA with soft CWIC 2 leads to almost optimal performance with low complexity. To save buffer and delay, CC rate oriented RSMA with soft SLIC and joint de-mapper become promising candidates. We again observe that RSMA is superior to SDMA even under with similar decoding complexity.

\section{Conclusion} 
To conclude, we consider the practical implementation of RSMA and have addressed several problems. We first introduced three low-complexity power allocation designs that respectively optimize sum-rate, CC sum-rate with SIC receiver and CC sum-rate without SIC receiver. We then introduced several transceiver implementations that can be applied with RSMA. We noticed that these designs can be classified into either SIC-type receivers or non-SIC-type receivers based on their principle and proposed to choose the power allocation design in accordance with this classification. Through numerical simulations, we reveal the fact that, under practical constellation constraints, sum-rate oriented designs do not always bring the benefit of RSMA, while CC sum-rate oriented designs better exploit the potential of RSMA. It was also observed that, with CC sum-rate oriented precoder design, RSMA preserves its superiority over SDMA even without SIC receivers. LLS was performed to verify this conclusion. Different practical receiver designs were also compared using LLS. We show that the most common SIC-type receiver design, the hard CWIC, is not promising due to its weak immunity to error propagation. Soft CWICs suffer very little from error propagation and hence provide better performance. Joint de-mapping and soft SLIC, which incur less complexity on the user side with acceptable performance loss, are also promising candidates.

\appendices 
\section{Constellation-constrained rates}
Consider a general MISO system with
\begin{equation}
    y = \mathbf{h}^H \mathbf{P} \mathbf{s} + n,
\end{equation}
where $\mathbf{s}$ belongs to a finite constellation set $\mathcal{X}$ and $|\mathcal{X}| = M$. Let $\mathbf{s}_m$ be the $m$-th element in $\mathcal{X}$.
Then the conditional entropy of $\mathbf{s}$ given $\mathbf{h}$, $\mathbf{P}$ is given by (\ref{con_entropy}), where $H()$ represents the entropy\cite{fin_const_1}.
\begin{figure*}[!h]
\hrule
\begin{flalign}\label{con_entropy}
    H(\mathbf{s}|y,\mathbf{h},\mathbf{P})
    &= \frac{1}{M} \sum_{m=1}^{M} \mathsf{E}_n \Big\{ \log_2 \sum_{l=1}^{M} \exp \Big( -\frac{|\mathbf{h}^H \mathbf{P} (\mathbf{s}_m - \mathbf{s}_l) + n|^2 - |n|^2}{\sigma^2} \Big) \Big\}&&\nonumber\\
    &= \frac{1}{\ln2} + \frac{1}{M} \sum_{m=1}^{M} \mathsf{E}_n \Big\{ \log_2 \sum_{l=1}^{M} \exp \Big( -\frac{|\mathbf{h}^H \mathbf{P} (\mathbf{s}_m - \mathbf{s}_l) + n|^2}{\sigma^2} \Big) \Big\}&&
\end{flalign}

\begin{flalign}\label{con_entropy_approx}
    H(\mathbf{s}|y,\mathbf{h},\mathbf{P}) \approx 
      \frac{1}{M} \sum_{m=1}^{M} \log_2 \sum_{l=1}^{M} \exp \Big( -\frac{|\mathbf{h}^H \mathbf{P} (\mathbf{s}_m - \mathbf{s}_l)|^2}{2\sigma^2} \Big)&&
\end{flalign}

\begin{flalign}\label{CC_rate_common_1}
    R_{\text{CC},\text{c},k}
    =& I(s_\text{c};y_k|\mathbf{h}_k,\mathbf{P})&&\nonumber\\
    =& I\Big(\{s_\text{c},s_1,...,s_K\} ; y_k \Big| \mathbf{h}_k,\mathbf{P}\Big) - I\Big(\{s_1,...,s_K\} ; y_k \Big| \mathbf{h}_k,\mathbf{P},s_\text{c}\Big)&&\nonumber\\
    =& H(s_\text{c},s_1,...,s_K) - H\Big(s_\text{c},s_1,...,s_K \Big| y_k,\mathbf{h}_k,\mathbf{P}\Big) - H(s_1,...,s_K) + H\Big(s_1,...,s_K \Big| y_k,\mathbf{h}_k,\mathbf{P},s_\text{c}\Big)&&
\end{flalign}

\begin{flalign}\label{CC_rate_common_2}
    R_{\text{CC},\text{c},k}
    =& H(s_\text{c},s_k) - H\Big(s_\text{c},s_k \Big| y_k,\mathbf{h}_k,[\mathbf{p}_\text{c},\mathbf{p}_k]\Big) - H(s_k) + H\Big(s_k \Big| y_k,\mathbf{h}_k,[\mathbf{p}_\text{c},\mathbf{p}_k],s_\text{c}\Big)&&
\end{flalign}
\hrule
\end{figure*}
By applying Jensen's inequality and adding a constant, a very good approximation of (\ref{con_entropy}) can be obtained as (\ref{con_entropy_approx}) \cite{fin_const_2}.

Now consider a system model using RSMA with $s_\text{c} \in \mathcal{X}_\text{c}$ and $s_k \in \mathcal{X}_k, k \in \{1,...,K\}$. The constellation-constrained achievable rate of the common stream at user-$k$ is expressed as (\ref{CC_rate_common_1}) where $I(;)$ represents the mutual information \cite{Cover}. Since the directions of $\{\mathbf{p}_1,...,\mathbf{p}_k\}$ are designed using ZF precoding, (\ref{CC_rate_common_1}) is simplified into (\ref{CC_rate_common_2}).

It is assumed that probability distribution of the constellation points are uniform, hence
\begin{equation}\label{entropy_1}
    H(s_\text{c},s_k) = \log_2 |\mathcal{X}_\text{c} \times \mathcal{X}_k|
\end{equation}
and
\begin{equation}\label{entropy_2}
    H(s_k) = \log_2 |\mathcal{X}_k|.\\
\end{equation}
Assuming user-$k$ successfully decodes $s_\text{c}$ and applies SIC before decoding $s_k$, the constellation-constrained achievable rate of $s_k$ is expressed as
\begin{equation}\label{CC_rate_private_SIC}
\begin{split}
    R_{\text{CC},k}^{\text{SIC}}
    &= I \Big(s_k ; y_k \Big| \mathbf{h}_k,[\mathbf{p}_\text{c},\mathbf{p}_k],s_\text{c} \Big)\\
    &= H(s_k) - H \Big(s_k \Big| y_k - \mathbf{h}_k^H\mathbf{p}_\text{c} s_\text{c},\mathbf{h}_k,\mathbf{p}_k \Big),
\end{split}
\end{equation}
where $|\mathbf{h}_k^H \mathbf{p}_i|^2 = 0, \forall i \neq k$ has been used. Alternatively, if user-$k$ does not apply SIC but treats $s_\text{c}$ as noise when decoding $s_k$,
\begin{equation}\label{CC_rate_private_no_SIC}
\begin{split}
    &R_{\text{CC},k}^{\text{non-SIC}}\\
    =& I \Big(s_k ; y_k \Big| \mathbf{h}_k, [\mathbf{p}_\text{c},\mathbf{p}_k] \Big)\\
    =& I \Big(\{s_\text{c}, s_k\} ; y_k \Big| \mathbf{h}_k, [\mathbf{p}_\text{c},\mathbf{p}_k] \Big) - I \Big(s_\text{c} ; y_k \Big| \mathbf{h}_k, [\mathbf{p}_\text{c},\mathbf{p}_k], s_k \Big)\\
    =& H(s_\text{c}, s_k) - H \Big(s_\text{c}, s_k \Big| y_k,\mathbf{h}_k, [\mathbf{p}_\text{c},\mathbf{p}_k] \Big)\\
     &- H(s_\text{c}) + H \Big(s_\text{c} \Big| y_k,\mathbf{h}_k, [\mathbf{p}_\text{c},\mathbf{p}_k], s_k \Big)
\end{split}
\end{equation}
The CC sum-rate of RSMA is expressed as
\begin{equation}\label{SR_SIC}
    R_{\text{CC},\text{sum}}^{\text{SIC}} = \min_{k \in \{1,...,K\}} R_{\text{CC},\text{c},k} + \sum_{k=1}^{K} R_{\text{CC},k}^{\text{SIC}}
\end{equation}
when SIC is applied at the receivers and 
\begin{equation}\label{SR_non_SIC}
    R_{\text{CC},\text{sum}}^{\text{non-SIC}} = \min_{k \in \{1,...,K\}} R_{\text{CC},\text{c},k} + \sum_{k=1}^{K} R_{\text{CC},k}^{\text{non-SIC}}
\end{equation}
when the receivers does not use SIC.

By combining (\ref{con_entropy}), (\ref{con_entropy_approx}), (\ref{CC_rate_common_2}-\ref{CC_rate_private_SIC}) and (\ref{SR_SIC}), Proposition \ref{prop_1} is proved. By combining (\ref{con_entropy}), (\ref{con_entropy_approx}), (\ref{CC_rate_common_2}-\ref{entropy_2}), (\ref{CC_rate_private_no_SIC}) and (\ref{SR_non_SIC}), Proposition \ref{prop_2} is proved.

\section{The equivalence of joint de-mapping and treating interference as noise: A detection perspective}
Consider a receiver observing a symbol, $x$, which is corrupted by interference, $u$, and Gaussian noise, $n$. $x$ is associated with a binary label sequence, $\{c_1,...,c_L\}$, representing the code bits. The received signal is expressed as
\begin{equation}
    y = x + u + n,
\end{equation}
where $x$ and $u$ are from finite constellation sets, $\mathcal{X}$ and $\mathcal{U}$, following certain prior probability distributions, $P(x)$ and $P(u)$. For clarity, channel effects on the desired symbol and interference are considered implicitly in $x$ and $u$. 

The receiver detects $x$ by treating $u$ as part of the noise. To do so, we treat the sum of interference and noise as an effective noise, $n' \triangleq u+n$, following a probability density function given by Gaussian mixture,
\begin{equation}
    f_{N'}(n') = \sum_{u\in\mathcal{U}} P(u) f_N(n'-u),
\end{equation}
where $f_N(n)$ is the probability density function of $n$.
The optimum soft-output maximum a posteriori (MAP) de-mapper for $x$ in the presence of $n'$ calculates the LLR of the $l$-th code bit, $c_l$, carried by $x$ according to
\begin{equation}\label{TIN_LLR}
\begin{split}
    \Lambda_l = &\log \frac{\sum_{x\in\mathcal{X}_l^0}P(y|x)P(x)}{\sum_{x\in\mathcal{X}_l^1}P(y|x)P(x)}\\
    = & \log \frac{\sum_{x\in\mathcal{X}_l^0}f_{N'}(y-x)P(x)}{\sum_{x\in\mathcal{X}_l^1}f_{N'}(y-x)P(x)}\\
    = &\log \frac{\sum_{x\in\mathcal{X}_l^0}\sum_{u\in\mathcal{U}}f_N(y-x-u)P(x)P(u)}{\sum_{x\in\mathcal{X}_l^1}\sum_{u\in\mathcal{U}}f_N(y-x-u)P(x)P(u)},
\end{split}
\end{equation}
where $\mathcal{X}_l^0$ and $\mathcal{X}_l^1$ are the set of symbols with the $c_l$-th bit being 0 or 1.

Note that (\ref{TIN_LLR}) is exactly the same expression to use if the receiver jointly de-maps $x$ and $u$. Hence, we conclude that jointly de-mapping the desired signal and interference is equivalent to the optimal detector for signals corrupted by finite-constellation interference and noise when interference is treated as noise.

\section{Channel realizations}
The channel realizations used to produce Fig. 12 are as follows.

Channel 1:
\begin{equation}
    [\mathbf{h}_1 \; \mathbf{h}_2]= \begin{bmatrix}
        -0.3010-1.0804j & -0.6889-0.3357j\\
        -0.3562+0.8451j & 0.7694+0.0572j\\
        0.5462-0.2465j & -1.0358-0.1752j\\
        -0.1919-0.1582j & 1.0553+0.5879j
    \end{bmatrix}
\end{equation}

Channel 2:
\begin{equation}
    [\mathbf{h}_1 \; \mathbf{h}_2]= \begin{bmatrix}
        0.9101+0.3530j & -0.0353+0.8155j\\
        -1.0081-0.7742j & 0.3403-0.6657j\\
        0.7555+1.2008j & -0.3133+0.4655j\\
        -0.2739-1.3710j & 0.1513-0.5489j
    \end{bmatrix}
\end{equation}

\section{Code rates used in LLS}
This appendix is to provide code rates used in the LLSs to generate Fig. {\ref{BER_performance}}-{\ref{SNR_complexity}}. The code rates for SDMA are uniform across all the users and can be computed based on the throughput given in captions. For example, in Fig. {\ref{BER_10}}, the throughput of 3 bits/symbol and QPSK leads to code rate of 0.75 for both users. The code rates for RSMA schemes are selected to reach good operating points, i.e. both common and private streams achieve $10^{-4}$ BER at similar SNRs. These code rates are summarized in Table. II in the format of "common stream code rate + private stream code rate" and, in all cases, all the private streams share the same code rate. In Fig. {\ref{SNR_complexity}}, SIC type and non-SIC type receivers use the same code rates as soft CWIC 1 and joint de-mapping respectively in Fig. {\ref{BER_10}}.

\begin{table*}
\caption{Code rates used for RSMA in link-level simulations.}
\centering
\begin{tabular}{|c|c|c|c|c|c|c|c|}
\hline
 & Fig. \ref{BER_10} & Fig. \ref{BER_10_16QAM} & Fig. \ref{BER_block_fading} & Fig. \ref{BER_three_user} & \multicolumn{3}{c|}{Fig. \ref{different_delta}, $\Delta=\pi/18, \pi/9, \pi/6$}\\
 \hline
 Soft CWIC 1 (finite) & 0.81 + 0.345 & 0.75 + 0.375 & 0.88 + 0.46 & 0.78 + 0.4 & 0.8 + 0.35& 0.7 + 0.4& 0.6 + 0.45\\
 \hline
 Joint de-mapping (finite) & 0.85 + 0.325 & 0.78 + 0.36 & 0.94 + 0.43 & 0.84 + 0.36 & - & - & -\\
 \hline
 Soft CWIC 1 (Gaussian) & 0.815 + 0.3425 & 0.7 + 0.4 & 0.8 + 0.5 & 0.66 + 0.48 & - & - & -\\
 \hline
 Joint de-mapping (Gaussian) & 0.82 + 0.34 & 0.75 + 0.375 & 0.8 + 0.5 & 0.69 + 0.46 & - & - & -\\
 \hline
\end{tabular}
\end{table*}

\bibliographystyle{IEEEtran}

\begin{thebibliography}{10}
\providecommand{\url}[1]{#1}
\csname url@samestyle\endcsname
\providecommand{\newblock}{\relax}
\providecommand{\bibinfo}[2]{#2}
\providecommand{\BIBentrySTDinterwordspacing}{\spaceskip=0pt\relax}
\providecommand{\BIBentryALTinterwordstretchfactor}{4}
\providecommand{\BIBentryALTinterwordspacing}{\spaceskip=\fontdimen2\font plus
\BIBentryALTinterwordstretchfactor\fontdimen3\font minus \fontdimen4\font\relax}
\providecommand{\BIBforeignlanguage}[2]{{%
\expandafter\ifx\csname l@#1\endcsname\relax
\typeout{** WARNING: IEEEtran.bst: No hyphenation pattern has been}%
\typeout{** loaded for the language `#1'. Using the pattern for}%
\typeout{** the default language instead.}%
\else
\language=\csname l@#1\endcsname
\fi
#2}}
\providecommand{\BIBdecl}{\relax}
\BIBdecl

\bibitem{Tataria}
H.~Tataria, M.~Shafi, A.~F. Molisch, M.~Dohler, H.~Sjöland, and F.~Tufvesson, ``{6G} wireless systems: Vision, requirements, challenges, insights, and opportunities,'' \emph{Proceedings of the IEEE}, vol. 109, no.~7, pp. 1166--1199, 2021.

\bibitem{Clerckx_2}
B.~Clerckx, Y.~Mao, E.~A. Jorswieck, J.~Yuan, D.~J. Love, E.~Erkip, and D.~Niyato, ``A primer on rate-splitting multiple access: Tutorial, myths, and frequently asked questions,'' \emph{IEEE Journal on Selected Areas in Communications}, pp. 1--1, 2023.

\bibitem{Mao_3}
Y.~Mao, O.~Dizdar, B.~Clerckx, R.~Schober, P.~Popovski, and H.~V. Poor, ``Rate-splitting multiple access: Fundamentals, survey, and future research trends,'' \emph{IEEE Communications Surveys \& Tutorials}, vol.~24, no.~4, pp. 2073--2126, 2022.

\bibitem{Carleial}
A.~Carleial, ``Interference channels,'' \emph{IEEE Transactions on Information Theory}, vol.~24, no.~1, pp. 60--70, 1978.

\bibitem{Han}
T.~Han and K.~Kobayashi, ``A new achievable rate region for the interference channel,'' \emph{IEEE Transactions on Information Theory}, vol.~27, no.~1, pp. 49--60, 1981.

\bibitem{Piovano}
E.~Piovano and B.~Clerckx, ``Optimal {DoF} region of the $k$ -user {MISO BC} with partial {CSIT},'' \emph{IEEE Communications Letters}, vol.~21, no.~11, pp. 2368--2371, 2017.

\bibitem{Hamdi}
H.~Joudeh and B.~Clerckx, ``Sum-rate maximization for linearly precoded downlink multiuser {MISO} systems with partial {CSIT}: A rate-splitting approach,'' \emph{IEEE Transactions on Communications}, vol.~64, no.~11, pp. 4847--4861, 2016.

\bibitem{Mao}
Y.~Mao and B.~Clerckx, ``Beyond dirty paper coding for multi-antenna broadcast channel with partial {CSIT}: A rate-splitting approach,'' \emph{IEEE Transactions on Communications}, vol.~68, no.~11, pp. 6775--6791, 2020.

\bibitem{Mao_2}
Y.~{Mao}, B.~{Clerckx}, and V.~O. {Li}, ``Rate-splitting multiple access for downlink communication systems: bridging, generalizing, and outperforming {SDMA} and {NOMA},'' \emph{EURASIP journal on wireless communications and networking}, vol. 2018, no.~1, p. 133, 2018.

\bibitem{Clerckx_1}
B.~Clerckx, Y.~Mao, R.~Schober, and H.~V. Poor, ``Rate-splitting unifying {SDMA, OMA, NOMA}, and multicasting in {MISO} broadcast channel: A simple two-user rate analysis,'' \emph{IEEE Wireless Communications Letters}, vol.~9, no.~3, pp. 349--353, 2020.

\bibitem{Dizdar_2}
O.~Dizdar, Y.~Mao, and B.~Clerckx, ``Rate-splitting multiple access to mitigate the curse of mobility in (massive) {MIMO} networks,'' \emph{IEEE Transactions on Communications}, vol.~69, no.~10, pp. 6765--6780, 2021.

\bibitem{Lu}
G.~Lu, L.~Li, H.~Tian, and F.~Qian, ``{MMSE}-based precoding for rate splitting systems with finite feedback,'' \emph{IEEE Communications Letters}, vol.~22, no.~3, pp. 642--645, 2018.

\bibitem{Hao}
C.~Hao, Y.~Wu, and B.~Clerckx, ``Rate analysis of two-receiver {MISO} broadcast channel with finite rate feedback: A rate-splitting approach,'' \emph{IEEE Transactions on Communications}, vol.~63, no.~9, pp. 3232--3246, 2015.

\bibitem{Flores}
A.~R. Flores, R.~C. de~Lamare, and B.~Clerckx, ``Linear precoding and stream combining for rate splitting in multiuser {MIMO} systems,'' \emph{IEEE Communications Letters}, vol.~24, no.~4, pp. 890--894, 2020.

\bibitem{Zhou}
G.~Zhou, Y.~Mao, and B.~Clerckx, ``Rate-splitting multiple access for multi-antenna downlink communication systems: Spectral and energy efficiency tradeoff,'' \emph{IEEE Transactions on Wireless Communications}, vol.~21, no.~7, pp. 4816--4828, 2022.

\bibitem{Dai}
M.~Dai and B.~Clerckx, ``Multiuser millimeter wave beamforming strategies with quantized and statistical {CSIT},'' \emph{IEEE Transactions on Wireless Communications}, vol.~16, no.~11, pp. 7025--7038, 2017.

\bibitem{Salem}
A.~Salem, C.~Masouros, and B.~Clerckx, ``Rate splitting with finite constellations: The benefits of interference exploitation vs suppression,'' \emph{IEEE Open Journal of the Communications Society}, vol.~2, pp. 1541--1557, 2021.

\bibitem{Dizdar}
O.~Dizdar, Y.~Mao, W.~Han, and B.~Clerckx, ``Rate-splitting multiple access for downlink multi-antenna communications: Physical layer design and link-level simulations,'' in \emph{2020 IEEE 31st Annual International Symposium on Personal, Indoor and Mobile Radio Communications}, 2020, pp. 1--6.

\bibitem{Yin}
L.~Yin, O.~Dizdar, and B.~Clerckx, ``Rate-splitting multiple access for multigroup multicast cellular and satellite communications: {PHY} layer design and link-level simulations,'' in \emph{2021 IEEE International Conference on Communications Workshops (ICC Workshops)}, 2021, pp. 1--6.

\bibitem{Chen_and_Mi}
H.~Chen, D.~Mi, Z.~Chu, P.~Xiao, Y.~Xu, and D.~He, ``Link-level performance of rate-splitting based downlink multiuser miso systems,'' in \emph{2020 IEEE 31st Annual International Symposium on Personal, Indoor and Mobile Radio Communications}, 2020, pp. 1--5.

\bibitem{Harshan}
J.~Harshan and B.~S. Rajan, ``On two-user gaussian multiple access channels with finite input constellations,'' \emph{IEEE Transactions on Information Theory}, vol.~57, no.~3, pp. 1299--1327, 2011.

\bibitem{Dong}
Z.~Dong, H.~Chen, J.-K. Zhang, L.~Huang, and B.~Vucetic, ``Uplink non-orthogonal multiple access with finite-alphabet inputs,'' \emph{IEEE Transactions on Wireless Communications}, vol.~17, no.~9, pp. 5743--5758, 2018.

\bibitem{fin_const_1}
Y.~Wu, M.~Wang, C.~Xiao, Z.~Ding, and X.~Gao, ``Linear precoding for {MIMO} broadcast channels with finite-alphabet constraints,'' \emph{IEEE Transactions on Wireless Communications}, vol.~11, no.~8, pp. 2906--2920, 2012.

\bibitem{Wu_2}
W.~Wu, K.~Wang, W.~Zeng, Z.~Ding, and C.~Xiao, ``Cooperative multi-cell {MIMO} downlink precoding with finite-alphabet inputs,'' \emph{IEEE Transactions on Communications}, vol.~63, no.~3, pp. 766--779, 2015.

\bibitem{Wu_3}
S.~X. Wu, Q.~Li, A.~M.-C. So, and W.-K. Ma, ``Rank-two beamforming and stochastic beamforming for {MISO} physical-layer multicasting with finite-alphabet inputs,'' \emph{IEEE Signal Processing Letters}, vol.~22, no.~10, pp. 1614--1618, 2015.

\bibitem{loli2022modelbased}
R.~C. Loli, O.~Dizdar, B.~Clerckx, and C.~Ling, ``Model-based deep learning receiver design for rate-splitting multiple access,'' \emph{arXiv preprint arXiv:2205.00849}, 2022.

\bibitem{joint_decoding}
Z.~Li, S.~Yang, and S.~Shamai, ``On linearly precoded rate splitting for gaussian mimo broadcast channels,'' \emph{IEEE Transactions on Information Theory}, vol.~67, no.~7, pp. 4693--4709, 2021.

\bibitem{joint_decoding_2}
Z.~Li, C.~Ye, Y.~Cui, S.~Yang, and S.~Shamai, ``Rate splitting for multi-antenna downlink: Precoder design and practical implementation,'' \emph{IEEE Journal on Selected Areas in Communications}, vol.~38, no.~8, pp. 1910--1924, 2020.

\bibitem{Sena}
A.~S. de~Sena, P.~H.~J. Nardelli, D.~B. da~Costa, P.~Popovski, C.~B. Papadias, and M.~Debbah, ``Dual-polarized massive {MIMO-RSMA} networks: Tackling imperfect {SIC},'' \emph{IEEE Transactions on Wireless Communications}, vol.~22, no.~5, pp. 3194--3215, 2023.

\bibitem{An}
J.~An, O.~Dizdar, B.~Clerckx, and S.~Wonjae, ``Rate-splitting multiple access for multi-antenna broadcast channel with imperfect {CSIT and CSIR},'' in \emph{2020 IEEE 31st Annual International Symposium on Personal, Indoor and Mobile Radio Communications}, 2020, pp. 1--7.

\bibitem{DVB}
{DVB}, ``Study mission reports: Commercial \& {Technical} reviews of {WiB},'' Jun. 2018.


\bibitem{NAICS}
``Study on network-assisted interference cancellation and suppression
  ({NAIC}) for {LTE} ({Release 12}),'' {3GPP TR 36.866}, Tech. Rep., Mar. 2014.


\bibitem{MUST}
``Study on downlink multiuser superposition transmission {(MUST)} for
  lte ({Release 13}),'' {3GPP TR 36.859}, Tech. Rep., Dec. 2015.

\bibitem{Multicast_1}
N.~Sidiropoulos, T.~Davidson, and Z.-Q. Luo, ``Transmit beamforming for physical-layer multicasting,'' \emph{IEEE Transactions on Signal Processing}, vol.~54, no.~6, pp. 2239--2251, 2006.

\bibitem{Multicast_2}
K.~Son and W.~Choi, ``Single group multicast beamformer design using active constraints,'' in \emph{2020 International Conference on Information and Communication Technology Convergence (ICTC)}, 2020, pp. 1364--1366.

\bibitem{Multicast_3}
\BIBentryALTinterwordspacing
C.-L. Hsiao, J.-C. Guey, W.-H. Sheen, and R.-J. Chen, ``A two-user approximation-based transmit beamforming for physical-layer multicasting in mobile cellular downlink systems,'' \emph{Journal of the Chinese Institute of Engineers}, vol.~38, no.~6, pp. 742--750, 2015. [Online]. Available: \url{https://doi.org/10.1080/02533839.2015.1016879}
\BIBentrySTDinterwordspacing

\bibitem{Soft_SIC}
W.-J. Choi, K.-W. Cheong, and J.~Cioffi, ``Iterative soft interference cancellation for multiple antenna systems,'' in \emph{2000 IEEE Wireless Communications and Networking Conference. Conference Record (Cat. No.00TH8540)}, vol.~1, 2000, pp. 304--309 vol.1.

\bibitem{BICM}
P.~Fertl, J.~Jalden, and G.~Matz, ``Performance assessment of {MIMO-BICM} demodulators based on mutual information,'' \emph{IEEE Transactions on Signal Processing}, vol.~60, no.~3, pp. 1366--1382, 2012.

\bibitem{channel}
D.-S. Shiu, G.~Foschini, M.~Gans, and J.~Kahn, ``Fading correlation and its effect on the capacity of multielement antenna systems,'' \emph{IEEE Transactions on Communications}, vol.~48, no.~3, pp. 502--513, 2000.


\bibitem{3GPP_Rx}
``{NR demodulation performance evolution (Release 18)},'' {3GPP TR 38.878}, Tech. Rep., Sep. 2023.

\bibitem{fin_const_2}
W.~Zeng, C.~Xiao, and J.~Lu, ``A low-complexity design of linear precoding for {MIMO} channels with finite-alphabet inputs,'' \emph{IEEE Wireless Communications Letters}, vol.~1, no.~1, pp. 38--41, 2012.

\bibitem{Cover}
T.~Cover and J.~Thomas, \emph{Elements of Information Theory}, 2nd~ed.\hskip 1em plus 0.5em minus 0.4em\relax New York: Wiley \& Sons, 2006.

\end{thebibliography}


\begin{IEEEbiography}[{\includegraphics[width=1in,height=1.25in,clip,keepaspectratio]{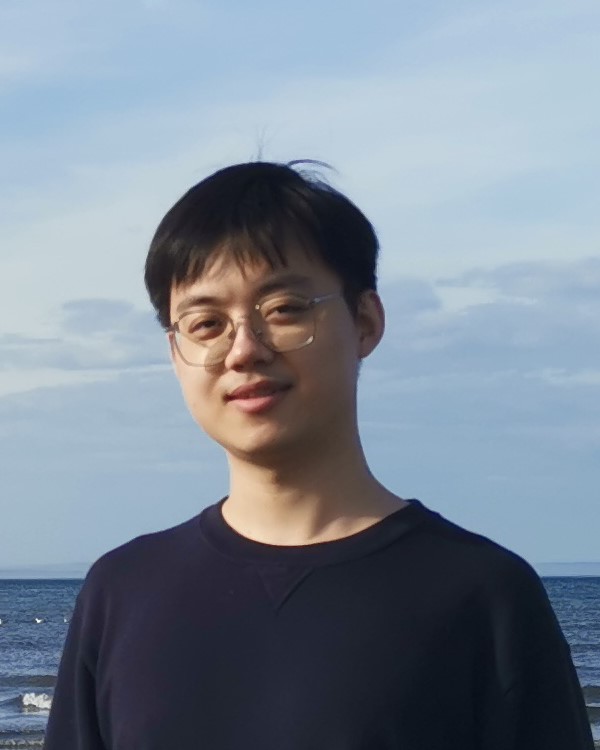}}]{Sibo Zhang}
received the B.Eng. degree in electrical and electronic engineering from University of Nottingham, UK, in 2019 and the M.S. degree in communications and signal processing from Imperial College London, UK, in 2021. He is currently pursuing his Ph.D. degree at Imperial College London, UK. He is also an intern at BBC Research \& Development in the distribution applied research area. His research interests include wireless communications, signal processing, information theory and optimization theory.
\end{IEEEbiography}

\begin{IEEEbiography}[{\includegraphics[width=1in,height=1.25in,clip,keepaspectratio]{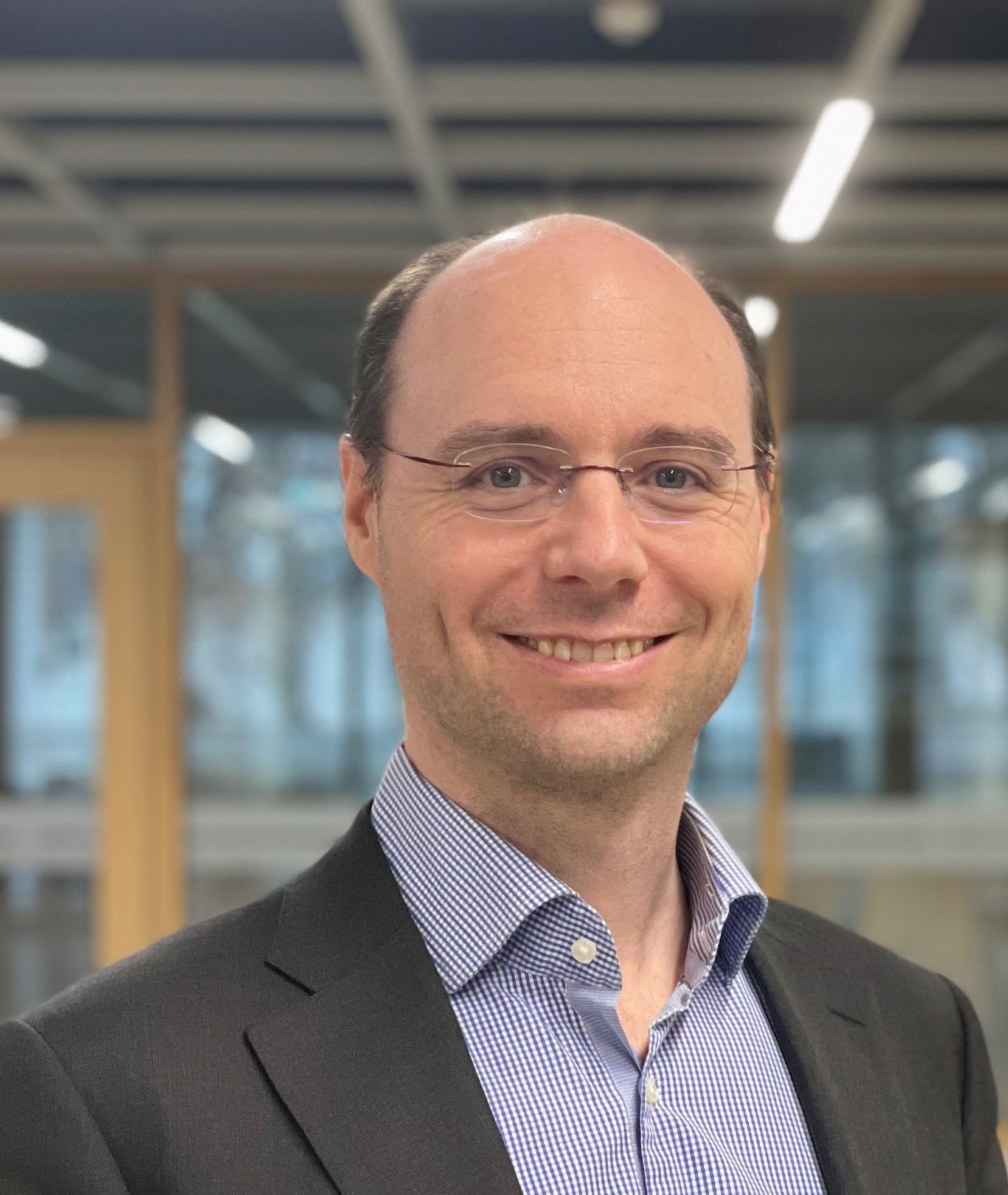}}]{Bruno Clerckx}
(Fellow, IEEE) is a Professor, the Head of the Wireless Communications and Signal Processing Lab, and the Head of the Communications and Signal Processing Group, within the Electrical and Electronic Engineering Department, Imperial College London, U.K.
\end{IEEEbiography}

\begin{IEEEbiography}[{\includegraphics[width=1in,height=1.25in,clip,keepaspectratio]{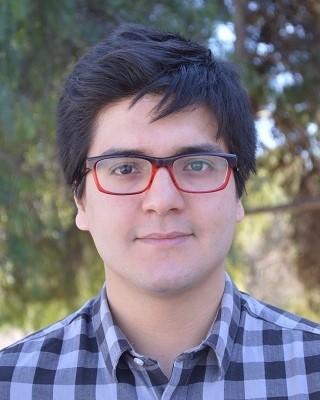}}]{David Vargas}
received the Ph.D. degree in Telecommunications from Universitat Politecnica de Valencia (UPV), Spain in 2016. From 2009 to 2015, he was a Research Engineer with the Institute of Telecommunications and Multimedia Applications, UPV. Since 2015, he has been with BBC R\&D, London, UK, where he is currently a Lead Research Engineer. David has contributed to standards development organisations such as DVB, ATSC, and 3GPP. He is currently Chair of the Content Distribution - Standards and Architectures working group at the 5G-MAG, a cross-industry association bringing together media and ICT industries collaborating on market-driven implementations of 5G. He has been a Guest Researcher with the Vienna University of Technology, and McGill University. His research interests include multi-antenna communications, signal processing for communications, and wireless communications.
\end{IEEEbiography}

\begin{IEEEbiography}[{\includegraphics[width=1in,height=1.25in,clip,keepaspectratio]{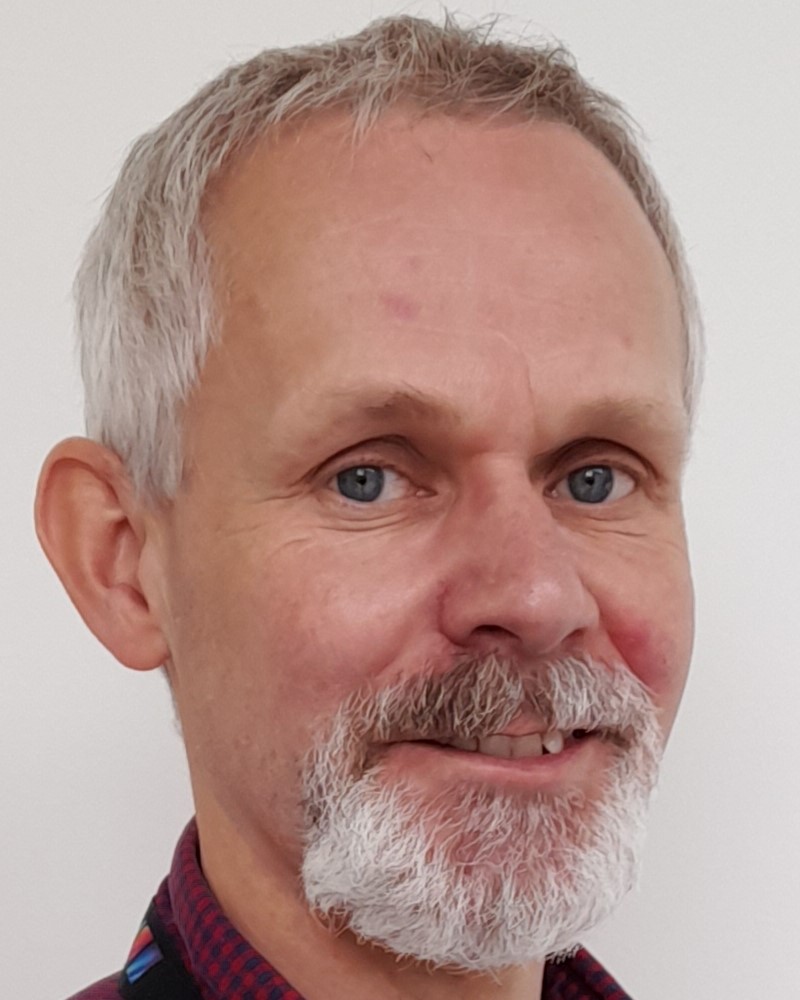}}]{Oliver Haffenden}
is a Lead R\&D Engineer at BBC Research \& Development. He has worked on standardisation and receiver algorithms for terrestrial modulation systems including DVB-T, DVB-T2 and Digital Radio Mondiale. He holds a Masters degree in Electrical and Information Sciences from the University of Cambridge.
\end{IEEEbiography}

\begin{IEEEbiography}[{\includegraphics[width=1in,height=1.25in,clip,keepaspectratio]{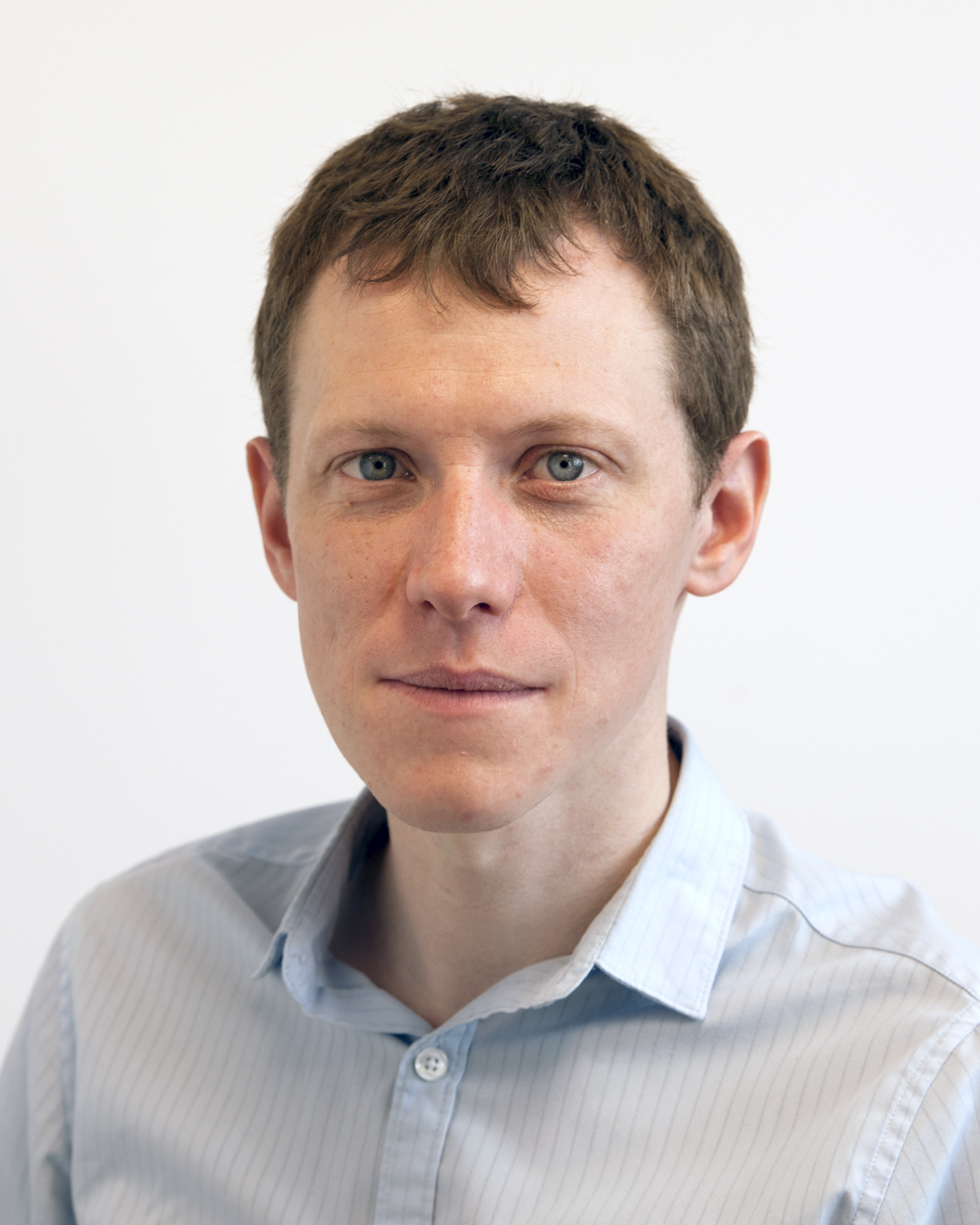}}]{Andrew Murphy}
is a Lead Research Engineer at BBC Research \& Development, where he leads work on distribution over mobile networks. He holds a Masters degree in Electrical and Information Sciences from the University of Cambridge.
\end{IEEEbiography}

\vfill

\end{document}